\documentclass[11pt,a4paper]{article} 
\pdfoutput=1
\usepackage{jheppub,hyperref,mdwlist}
\usepackage{amsmath}
\usepackage{graphicx}
\usepackage{subfigure}
\usepackage{amssymb}
\usepackage{xcolor}
\usepackage{multirow}
\usepackage{cancel}
\usepackage{color}
\usepackage{listings}
\usepackage{xcolor}
\usepackage{float}
\usepackage{slashed}

\usepackage{amsmath}
\usepackage{wrapfig}
\usepackage{cleveref}

\newcommand{\be}{\begin{equation}}
\newcommand{\ee}{\end{equation}}
\newcommand{\beq}{\begin{equation}}
\newcommand{\eeq}{\end{equation}}
\newcommand{\bea}{\begin{eqnarray}}
\newcommand{\eea}{\end{eqnarray}}
\newcommand{\besp}{\begin{equation}\begin{split}}
\newcommand{\eesp}{\end{split}\end{equation}}
\newcommand{\met}{\slashed{E}_{T}}
\newcommand{\nn}{\nonumber}

\newcommand{\Br}{\text{Br}}

\newcommand{\Eq}[1]{Eq.~(\ref{#1})}

\newcommand{\Dfbd}{\mathord{\buildrel{\lower3pt\hbox{$\scriptscriptstyle\leftrightarrow$}}\over {D}_{\mu}}}
\newcommand{\ave}[1]{\left\langle #1\right\rangle}

\def\mL{\mathcal{L}}

\def\mO{\mathcal{O}}

\def\mT{\mathcal{T}}

\def\Z{\mathbb{Z}}

\def\0{\textbf{0}}
\def\1{\textbf{1}}
\def\2{\textbf{2}}
\def\3{\textbf{3}}
\def\4{\textbf{4}}
\def\5{\textbf{5}}
\def\6{\textbf{6}}
\def\7{\textbf{7}}
\def\8{\textbf{8}}
\def\9{\textbf{9}}

\begin{document}

\title{Searching for lepton portal dark matter with colliders and gravitational waves}

\author[a,b]{Jia Liu,}
\author[c,d]{Xiao-Ping Wang,}
\author[e]{Ke-Pan Xie}
\affiliation[a]{School of Physics and State Key Laboratory of Nuclear Physics and Technology, Peking University, Beijing 100871, China}
\affiliation[b]{Center for High Energy Physics, Peking University, Beijing 100871, China}
\affiliation[c]{School of Physics, Beihang University, Beijing 100083, China}
\affiliation[d]{Beijing Key Laboratory of Advanced Nuclear Materials and Physics, Beihang University, Beijing 100191, China}
\affiliation[e]{Center for Theoretical Physics, Department of Physics and Astronomy, Seoul National University, Seoul 08826, Korea}

\emailAdd{jialiu@pku.edu.cn}
\emailAdd{hcwangxiaoping@buaa.edu.cn}
\emailAdd{kpxie@snu.ac.kr}

\abstract{

We study the lepton portal dark matter (DM) model in which the relic abundance is determined by the portal coupling among the Majorana fermion DM candidate $\chi$, the singlet charged scalar mediator $S^\pm$ and the Standard Model (SM) right-handed lepton. The direct and indirect searches are not sensitive to this model. This article studies the lepton portal coupling as well as the scalar portal coupling (between $S^\pm$ and SM Higgs boson), as the latter is generally allowed in the Lagrangian. The inclusion of scalar portal coupling not only significantly enhances the LHC reach via the $gg\to h^*\to S^+S^-$ process, but also provides a few novel signal channels, such as the exotic decays and coupling deviations of the Higgs boson, offering new opportunities to probe the model. In addition, we also study the Drell-Yan production of $S^+S^-$ at future lepton colliders, and find out that the scenario where one $S^\pm$ is off-shell can be used to measure the lepton portal coupling directly. In particular, we are interested in the possibility that the scalar potential triggers a first-order phase transition and hence provides the stochastic gravitational wave (GW) signals. In this case, the terrestrial collider experiments and space-based GW detectors serve as complementary approaches to probe the model.

}

\maketitle
\flushbottom

\section{Introduction}

The Standard Model (SM) of particle physics has been a great triumph in explaining and predicting the astrophysical and terrestrial experimental phenomena, however there are still many unsolved problems remaining, such as the dark matter (DM). Many astrophysical evidences support the existence of DM, and the fitting result of Cosmic Microwave Background (CMB) to the $\Lambda$CDM model yields a DM relic abundance of $\Omega_{\rm DM}h^2\approx0.12$~\cite{Aghanim:2018eyx}, which accounts for $\sim27\%$ of the total universe energy. However, we still know very little about the particle origin of DM, except that none of the SM particles can be the DM candidate~\cite{Bertone:2004pz}. Therefore, the existence of DM is a clear evidence for physics beyond the SM (BSM).

Over the past several decades, the most popular particle explanation for DM has been the freeze-out mechanism of the weakly interacting massive particles (WIMPs)~\cite{Lee:1977ua}, as it naturally yields the observed DM relic density when the coupling of DM to the SM particles is of the order of the electroweak (EW) gauge couplings, and the DM mass is $\mO(100~{\rm GeV})$. Although the results from direct detection~\cite{Schumann:2019eaa}, indirect detection~\cite{Gaskins:2016cha} and collider searches~\cite{Boveia:2018yeb} have been pushing more and more stringent bounds on WIMPs, there is still room for this scenario. 
There are many simplified models~\cite{Alves:2011wf, Abdallah:2014hon, Abdallah:2015ter, Abercrombie:2015wmb} describe the interactions between DM and SM particles. One category of them couples DM to SM fermions through Yukawa interaction~\cite{Boehm:2003hm}, which is similar to neutralino-sfermion-fermion vertex in supersymmetric (SUSY) models. The interactions can induce $t$-channel annihilation diagrams for the DM pair. Such colored~\cite{Goodman:2011jq, Garny:2012eb, Liu:2013gba, An:2013xka, DiFranzo:2013vra, Bai:2013iqa, Papucci:2014iwa, deSimone:2014pda, Abdallah:2014hon,Garny:2014waa, Gomez:2014lva, Baker:2015qna} and uncolored~\cite{Liu:2013gba, Bai:2014osa, Chang:2014tea, Agrawal:2014ufa, Garny:2015wea} mediators have been studied in literature.

The lepton portal DM model is proposed in Ref.~\cite{Bai:2014osa}, which assumes a portal coupling among the SM lepton and two dark sector particles $S$ (scalar) and $\chi$ (fermion), where ``dark sector'' means an odd $\Z_2$ symmetry is assigned to $S$ and $\chi$. Depending on the mass hierarchy, the DM candidate could be $S$ or $\chi$, while the fermion DM case can be further classified into Dirac or Majorana DM scenarios. This particular model has been further studied in Refs.~\cite{Yu:2014mfa, Altmannshofer:2014cla, Yu:2014pra, Garny:2015wea, Agrawal:2015tfa, Ibarra:2015fqa, Cai:2015zza, Baek:2015fma, Chen:2015jkt, Berlin:2015njh, Agrawal:2015kje,  Mukherjee:2015axj, Evans:2016zau, Chao:2016lqd, Borah:2017dqx, Kowalska:2017iqv, Duan:2017pkq,  Yuan:2017ysv, Tang:2017lfb, Ge:2017tkd, Ding:2017jdr, Baker:2018uox, Hisano:2018bpz, Gaviria:2018cwb, Kavanagh:2018xeh, Kawamura:2020qxo, Okada:2020oxh, Ge:2020tdh, Boehm:2020wbt, Okawa:2020jea, Kowalska:2020zve, Verma:2021koo, Alvarado:2021fbw, Horigome:2021qof, Bai:2021bau,Jueid:2020yfj,Arcadi:2021glq,Arcadi:2021cwg,Calibbi:2018rzv}.
In this article, we consider the model with right-handed lepton portal, taking $\chi$ as Majorana DM candidate and 
$S$ as the charged scalar mediator, which is similar to the setup in Ref.~\cite{Bai:2014osa}.

This model has very small indirect search cross sections due to helicity suppression. Because the DM candidate only couples 
to charged leptons, its nuclear recoil cross section comes from loop diagrams. Thus, the direct search signal is also suppressed. Therefore, collider experiments are crucial in probing this model. The typical collider signal is the Drell-Yan pair production of the $S^\pm$ mediator, and its subsequent decay to $\chi$ and a charged lepton. In this article, we study this channel at the LHC and future $e^+e^-$ colliders, and in the latter case we include the off-shell $S$ pair production $S^\pm S^{\mp(*)}$, which provides a direct probe for the lepton portal coupling.

Different from previous studies, in addition to the lepton portal coupling, we also consider the Higgs portal coupling $|S|^2|H|^2$, which is in general allowed in the Lagrangian. The inclusion of this coupling leads to several novel signals, such as the gluon-gluon fusion production of $S^+S^-$ at the LHC, the Higgs exotic decay (e.g. $h\to\chi\chi$, $h\to\ell^+\ell^-\chi\chi$), the Higgs coupling (e.g. $hZZ$, $h\gamma\gamma$, $h\ell^+\ell^-$) deviations and the lepton $(g-2)$ corrections to the SM prediction. In particular, the scalar portal coupling might be able to trigger a first-order phase transition (FOPT) in the early universe, opening the window for detecting the model via the stochastic gravitational wave (GW) signals. We finally consider the interplay between the GW searches and collider searches and show their complementarity.

This paper is organized as follows. We describe the model and derive the parameter space for the WIMP DM candidate in section~\ref{sec:model}. In section~\ref{sec:collider}, various terrestrial searches are investigated, including the $S^+S^-$ production, the exotic decay and coupling deviations of the Higgs boson at the LHC and future $e^+e^-$ colliders, together with lepton $(g-2)$ searches to constrain the two portal couplings in this model. The scenario that this model provides the right DM relic abundance and at the same time triggers a FOPT is considered in section~\ref{sec:psgw}, in which we show that detectable GW signals suggest a large Higgs portal coupling. In section~\ref{sec:interplay}, the interplay between GW detectors and the collider experiments are discussed. Finally, we summarize and conclude in section~\ref{sec:conclusion}.

\section{The model}\label{sec:model}

The model contains two new fields: the Majorana DM candidate $\chi$, which is a gauge singlet; and the complex scalar mediator $S$, which is an $SU(2)_L$ singlet with hypercharge $-1$. The relevant Lagrangian reads
\begin{align}\label{eq:Lags}
\mL_\chi =&~ \frac12\bar\chi i\slashed{\partial}\chi-\frac12m_\chi\bar\chi\chi+ y_\ell\left(\bar{\chi}_L S^\dagger \ell_R+{\rm h.c.}\right), \\
\mL_S =&~ \left(D^\mu S \right)^\dagger D_\mu S - V(H,S),\\
V(H, S)=&~\mu_H^2|H|^2+\mu_S ^2| S |^2+\lambda_H|H|^4+\lambda_S | S |^4+2\lambda_{H S }|H|^2| S |^2,
\end{align}
where $H$ is the SM Higgs doublet, and $\ell=e,\mu,\tau$ is the SM charged lepton (mass eigenstate). We require $S$ couple to one generation of lepton at a time to avoid lepton flavor violation. Such a flavor alignment typically needs some specific underlying mechanism to realize, which however beyond the scope of this work. For a more generic model that simultaneously involves all three flavors, our results still apply, as long as suitable rescaling is performed. The model contains a $Z_2$ symmetry for $\chi$ and $S$, that both of them carry odd charges. Assuming $m_\chi < m_S $, the $\chi$ is stable and thus can be the DM candidate.

\begin{figure}
	\centering	
	\includegraphics[width=0.4\columnwidth]{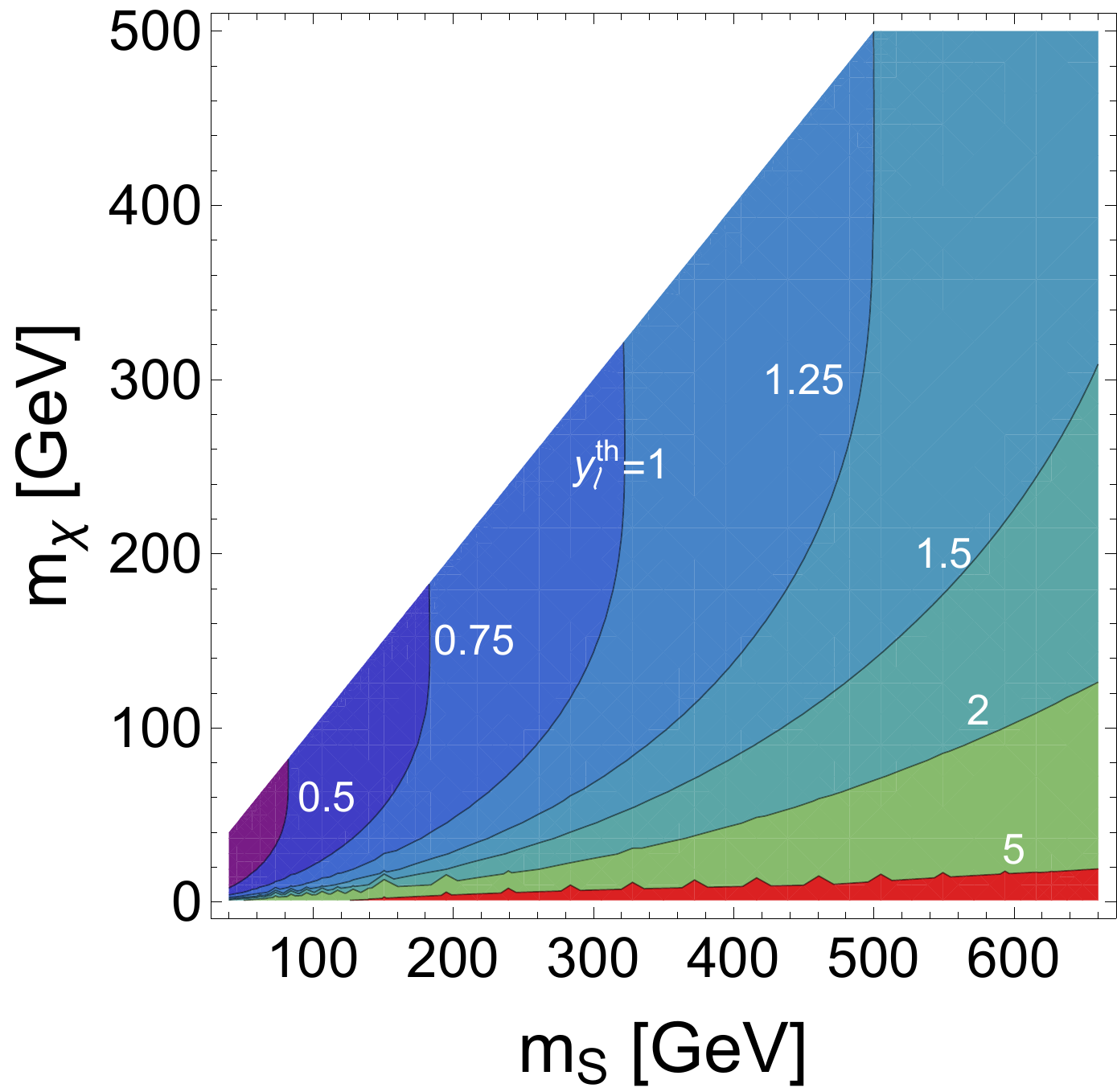}
	\caption{The lepton portal coupling $y_\ell^{\rm th}$ as a function of $m_S$ and $m_\chi$, which satisfies the DM
		relic abundance requirement.}
	\label{fig:DMrelic}
\end{figure}

The $\chi$ pair can annihilate into the lepton pair via the exchange of a $t$-channel $S$. Due to the Majorana nature of $\chi$, the $s$-wave component of $\chi\chi\to\ell^+\ell^-$ is suppressed by the lepton mass. Therefore, the annihilation cross section is $p$-wave dominated~\cite{Garny:2013ama, Luo:2013bua},
\be\label{eq:MajDM}
\sigma v_{\rm rel} = \frac{y^4_\ell }{32 \pi} \frac{m_\ell^2}{m_S^4} \frac{1}{(1+x)^2}
+ v_{\rm rel}^2 \frac{y^4_\ell }{48 \pi m_S^2} \frac{x(1+x^2)}{(1+x)^4} 
\approx  v_{\rm rel}^2 \frac{y^4_\ell }{48 \pi m_S^2} \frac{x(1+x^2)}{(1+x)^4},
\ee
where $x \equiv m_\chi^2/m_S^2$, and we have applied the limit $m_\ell \to 0$ in the second equality. We perform the thermal average of the annihilation cross section according to Ref.~\cite{Gondolo:1990dk}. Given a set of $(m_S,m_\chi)$, one can always tune $y_\ell$ to have the right annihilation cross section at freeze-out to achieve the observed DM relic abundance, and the corresponding $y_\ell$ is denoted as $y_\ell^{\rm th}$, which is plotted in Fig.~\ref{fig:DMrelic}. One can see that for EW scale $m_\chi$ and $m_S $, a Yukawa coupling $y_\ell\sim\mO(1)$ can provide the correct DM relic density. For $m_\chi < m_S$, smaller $m_\chi$ leads to larger $y_\ell^{\rm th}$, because the annihilation cross section scales as $m_\chi^2/m_S^4$.

As the DM annihilation signal $\chi\chi\to\ell^+\ell^-$ is helicity or $p$-wave suppressed, it is hard to be probed by satellite experiments like Fermi-LAT~\cite{Ackermann:2015zua, Fermi-LAT:2016uux}, AMS-02~\cite{Accardo:2014lma, Aguilar:2014mma}, or the CMB measurements from Planck \cite{Aghanim:2018eyx}. For direct detection, the scattering between $\chi$ and nucleons arises only at one-loop level, which can be described by an effective operator. Since $\chi$ is a Majorana fermion, its dimension-five magnetic dipole operator vanishes. It leaves the dimension-six operator as the leading contribution, which can be matched to the electromagnetic anapole moment of DM. This receives additional suppression from DM velocity square, so that it is difficult to detect from the direct detection experiments~\cite{Bai:2014osa}.\footnote{The low energy electron recoil cross section is $(y_\ell^4/\pi)m_e^2/(m_\phi^2-m_\chi^2)^2$, which is typically $10^{-44}~{\rm cm}^2$ for $m_\phi,~m_\chi\sim\mO(100)~{\rm GeV}$, well below the constraint from LUX-ZEPLIN experiment~\cite{Akerib:2021qbs}.} As a result, we conclude that the lepton portal DM with Majorana DM
has negligible signal in indirect and direct searches, and is only subject to the constraints from the thermal relic abundance and collider searches.

\section{The particle experiment searches}\label{sec:collider}

In this section, we discuss probing the lepton portal coupling $y_\ell$ and Higgs portal coupling $\lambda_{HS}$ in particle experiments. First, we consider the $S^+S^-\to\ell^+\chi\ell^-\chi$ channel at the LHC and future lepton colliders. In the latter case, the off-shell production of $S^\pm$ offers the opportunity to probe $y_\ell$ directly. Next, we study the exotic decays of the Higgs and $Z$ bosons, including the three/four-body decays $h/Z \to S^{\pm(*)} S^{\mp(*)}$ and the invisible decay $h\to\chi\chi$ at the Higgs factory CEPC and FCC-ee, which probe the combination of couplings $y_\ell$ and $\lambda_{HS}$. Then we turn to the correction to the Higgs couplings $h\ell^+\ell^-$, $h\gamma\gamma$ and $hZZ$. Finally, the lepton $(g-2)$ is discussed.

\subsection{Pair production of $S^\pm$}

\subsubsection{$pp\to S^+S^-$ at the LHC }\label{sec:pp_SS}

\begin{figure}
\centering
\includegraphics[width=0.418 \columnwidth]{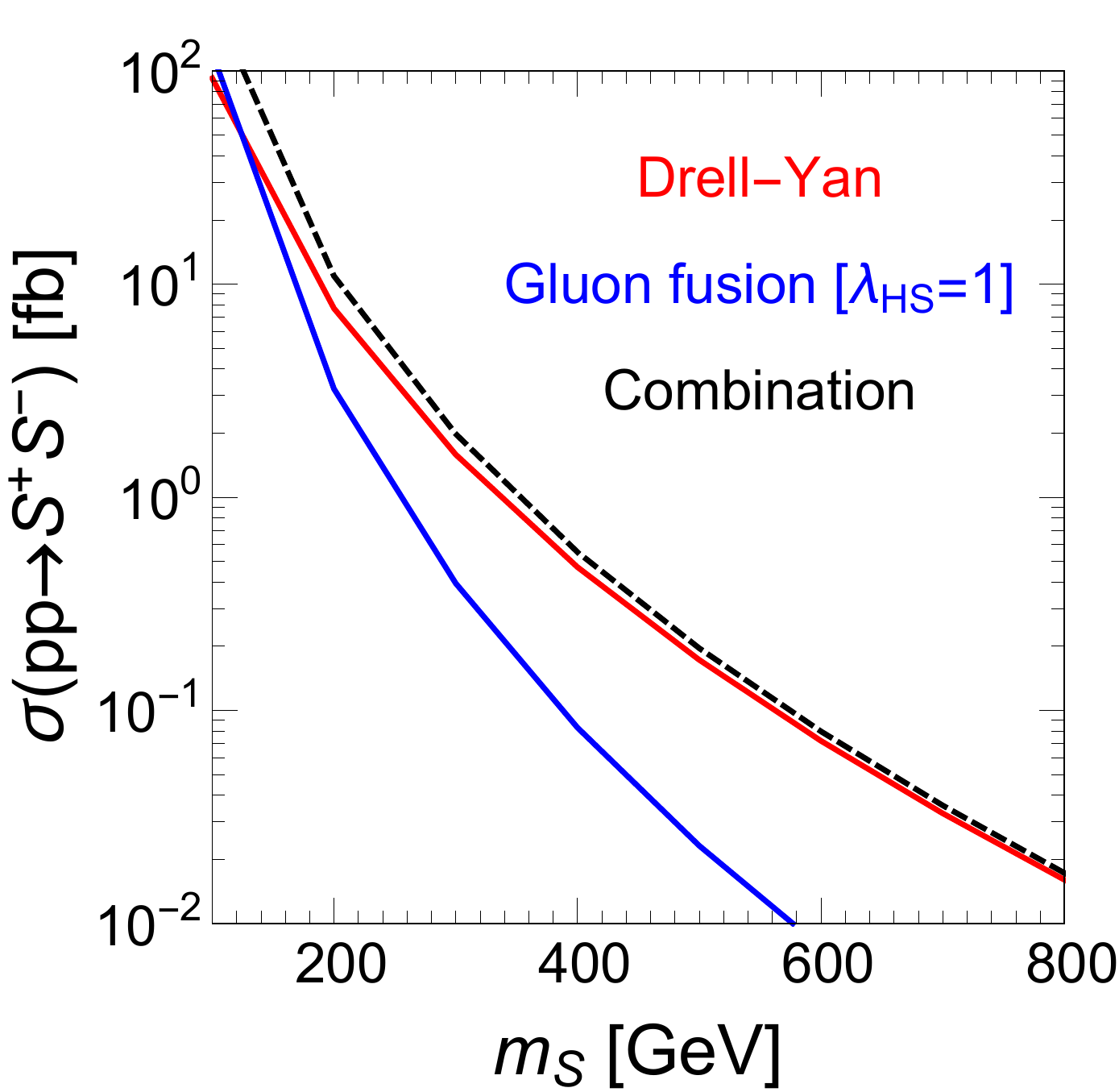}\qquad\qquad
\includegraphics[width=0.4 \columnwidth]{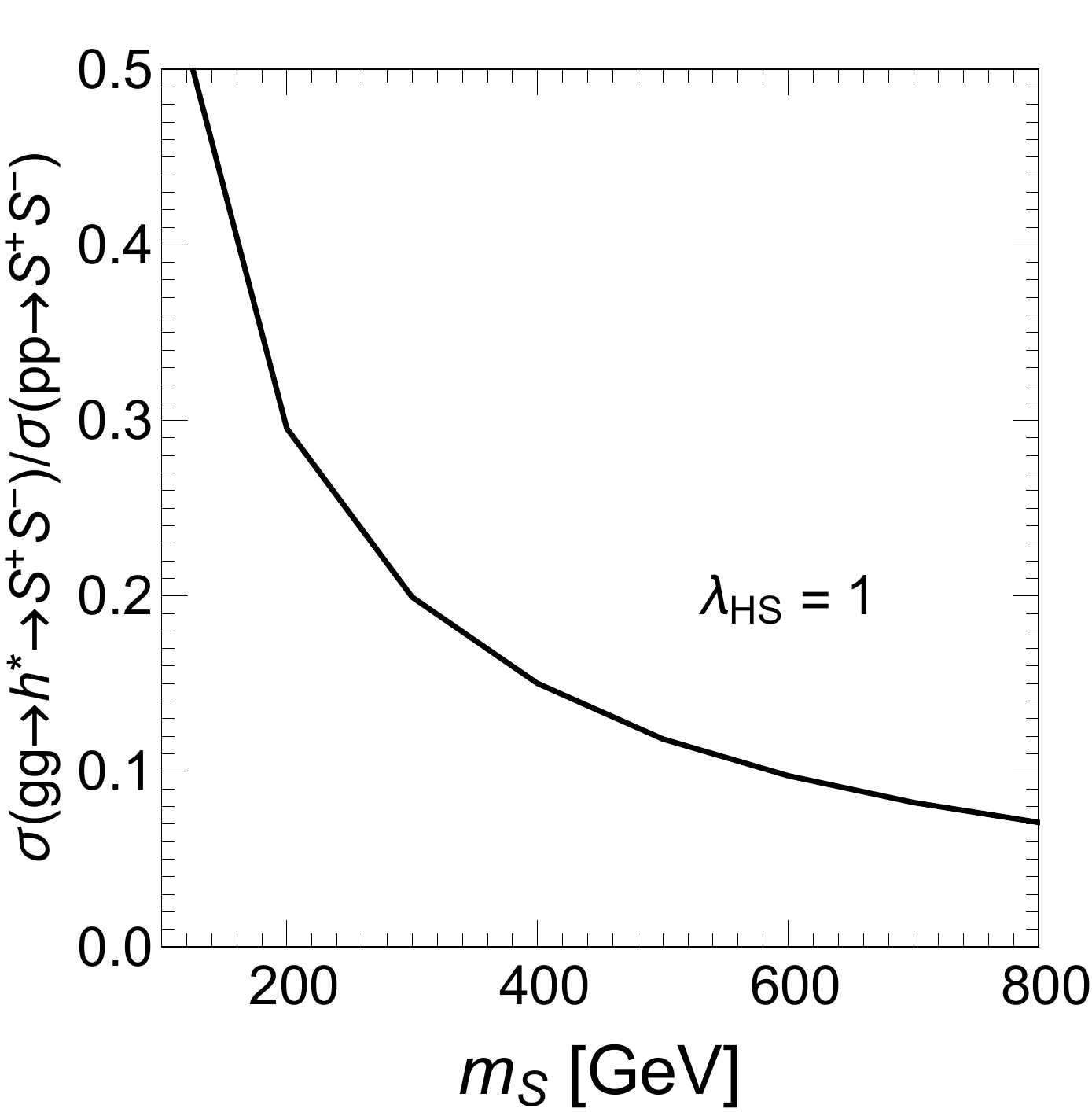}
\caption{Left: the production rate of $S^+S^-$ at the 14 TeV LHC for the Drell-Yan (red), gluon-gluon fusion (blue) channels and their combination (black dashed). Right: the ratio of gluon-gluon fusion rate to the total cross section, for $\lambda_{HS}=1$.}
\label{fig:SS_LHC}
\end{figure}

In the model, the lepton portal scalar $S$ carries one unit of hypercharge, therefore it can be produced 
in pair via the EW Drell-Yan process $q\bar q\to Z^*/\gamma^*\to S^+S^-$ mediated by off-shell $\gamma$ and $Z$ bosons. However, in our model due to the large scalar sector coupling $\lambda_{HS}$, one can also have the (off-shell) Higgs mediated process $p p \to h^{*} \to S^+ S^-$ which can significantly modify the total cross section of $pp \to S^+ S^-$. For example, with $\lambda_{HS} = 1$ and $m_S = 200$ GeV, the cross section contributed by the Higgs mediation can be $30\%$ of the total cross section. The production rates of the Drell-Yan and the gluon-gluon fusion processes are shown in Fig.~\ref{fig:SS_LHC}.

\begin{figure}
\centering	
\includegraphics[width=0.4 \columnwidth]{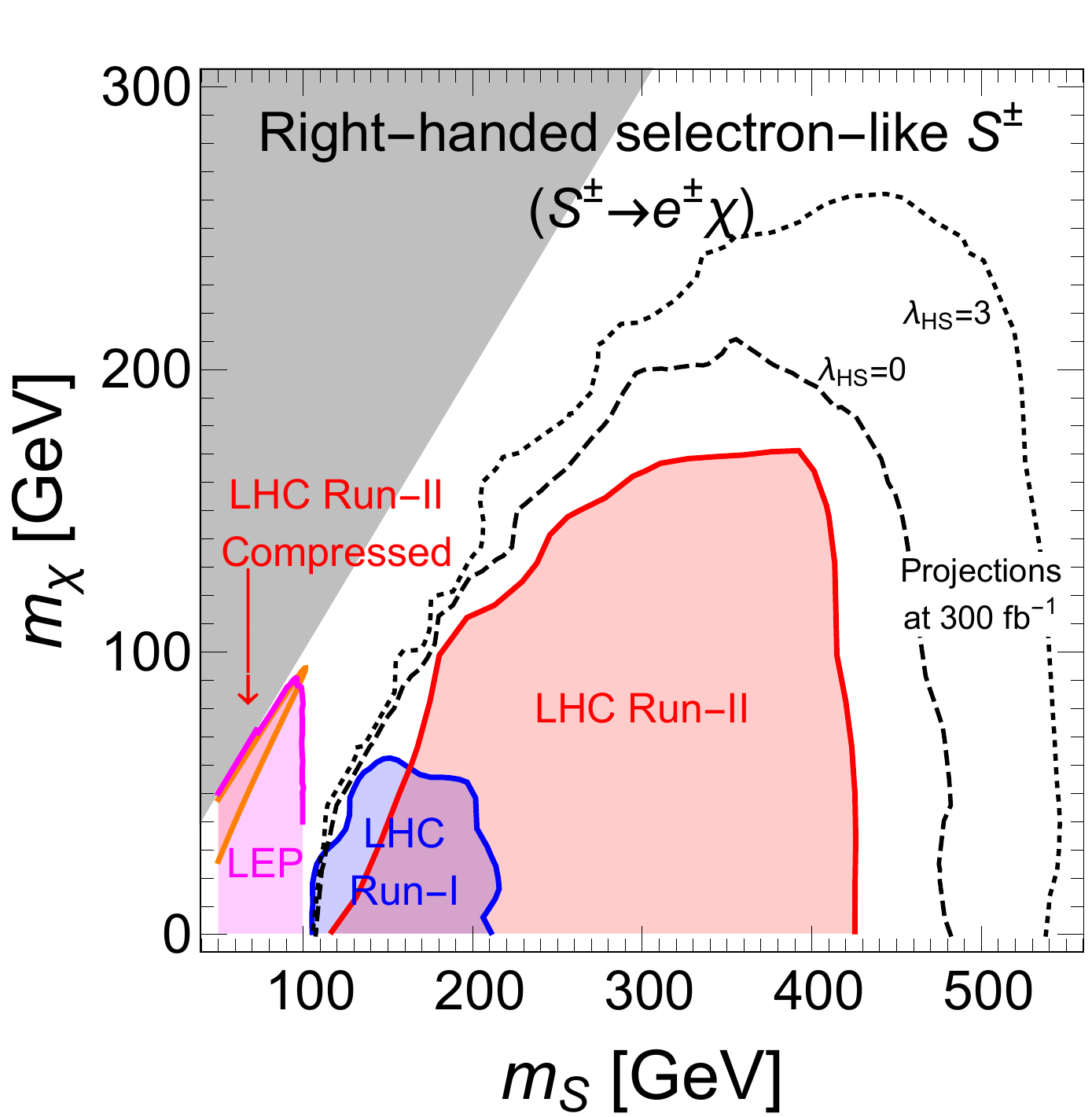}\qquad\qquad
\includegraphics[width=0.4 \columnwidth]{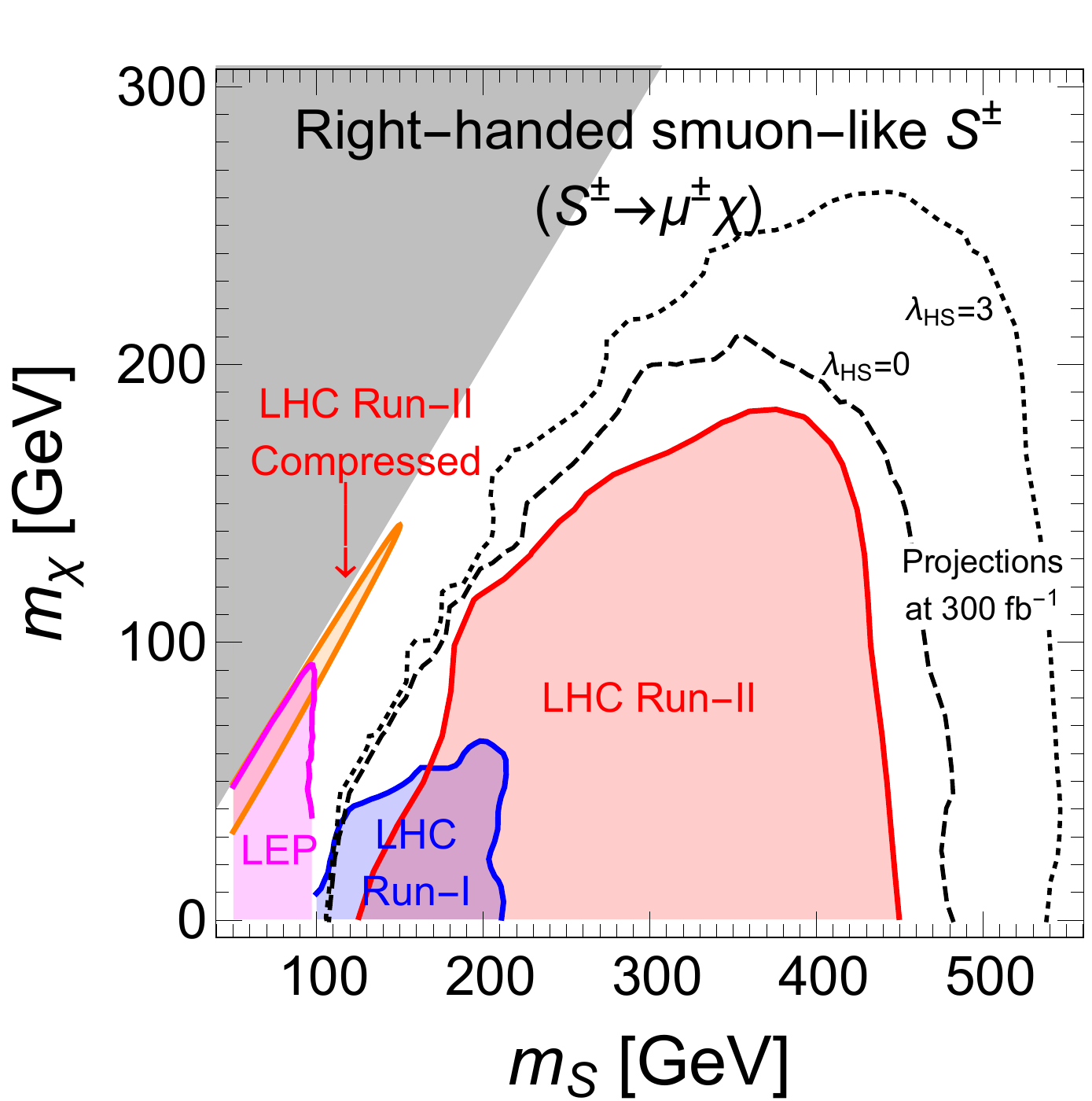}
\caption{The current constraints for right-handed selectron-like and right-handed smuon-like scalar $S$ in the $m_\chi$-$m_S$ plane. The shaded regions are exclusions from LEP~\cite{LEPslepton}, LHC Run-I ($20.3~{\rm fb}^{-1}$)~\cite{Aad:2014vma} and LHC Run-II ($139~{\rm fb}^{-1}$)~\cite{Aad:2019vnb,Aad:2019qnd}.	The black lines are projections for the LHC reach at 300 ${\rm fb}^{-1}$ with $\lambda_{HS}=0$ (solid) and $\lambda_{HS}=3$ (dashed).}\label{fig:slepton-constraints}
\end{figure}

The produced $S^\pm$ exclusively decays to $\ell^\pm \chi$, leading to a di-lepton plus missing transverse energy final state ($\ell^+\ell^-+\met$) at the LHC. The LHC experiments have already set constraints on such a final state in searches for the sleptons from SUSY models~\cite{Aad:2014vma, Aad:2019vnb, Aad:2019qnd}. The LHC Run-I and Run-II data from ATLAS \cite{Aad:2014vma, Aad:2019vnb} have covered mass of $S^\pm$ up to 450 GeV for the exclusive decay channel $e^\pm\chi$ or $\mu^\pm\chi$. The compressed parameter region when $m_\chi$ is close to $m_S$, has also been studied by the ATLAS collaboration~\cite{Aad:2019qnd}. Earlier studies from LEP have fully excludes such charged scalar $S$ with mass $m_S < 100$ GeV~\cite{LEPslepton}. We show the above existing constraints in Fig.~\ref{fig:slepton-constraints} as colored regions.

To make future projections for our model, we write down the UFO model file~\cite{Degrande:2011ua} with the {\tt FeynRules} package~\cite{Alloul:2013bka} and use {\tt MadGraph5\_aMC@NLO}~\cite{Alwall:2014hca} to generate parton level events, and then use {\tt Pythia8}~\cite{Sjostrand:2007gs} and {\tt Delphes}~\cite{deFavereau:2013fsa} to implement parton shower and fast detector simulation, respectively. Both the EW Drell-Yan and the Higgs mediated $S^+S^-$ production processes are included in our simulation. The signal events are selected using the cuts from two signal regions (SRs) of the ATLAS study~\cite{Aad:2019vnb} as follows,
\begin{enumerate}
\item Exactly two opposite charged leptons with $p_T>25$ GeV and $|\eta|<2.47$; 
\item At most one light-flavor jet with $p_T>20$ GeV and $|\eta|<2.4$, and veto the $b$-jets in such kinematic region;
\item The invariant mass $m_{\ell\ell}>100$ GeV, and transverse missing momentum $\met>110$ GeV;
\item $m_{T2}>100$ or 160 GeV.
\end{enumerate}
Here the $m_{T2}$ observable is defined event-by-event as the minimum of the function~\cite{Lester:1999tx}
\be
\max\left\{ m_T(\vec{p}_T^{\,\ell^+},\vec{p}_T^{\,a}),m_T(\vec{p}_T^{\,\ell^-},\vec{p}_T^{\,b})\right\} ,
\ee
subject to $\vec{p}^{\,a}+  \vec{p}^{\,b}= \met$, where $\vec{p}_T^{\,\ell^\pm}$ are the transverse momentum of the two charged leptons, $\vec{p}^{\,a}$ and $\vec{p}^{\,b}$ are the associated missing momenta. The transverse mass $m_{T}$ is defined as 
\be
m_{T}(\vec{p}_{T,1}, \vec{p}_{T,2}) = \sqrt{|\vec{p}_{T,1}||\vec{p}_{T,2}|(1-\cos\Delta\phi_{12}) }.
\ee
The $m_{T2}$ cut significantly suppresses the $W^+W^-$ and $t\bar t$ backgrounds as their $m_{T2}$ have an end point at the $m_W=80.4$ GeV.

According to the number (1 or 2) of light-flavor jets in the final state and the different $m_{T2}$ cuts (100 or 120 GeV), we classify the events into 4 SRs, and adopt the simulated background event numbers from Ref.~\cite{Aad:2019vnb}. We use $S/\sqrt{B}=2$ as the criterion of the LHC reach for a given integrated luminosity, where $S$ and $B$ denote the signal and background event numbers, respectively. The projections at the 14 TeV LHC with 300 fb$^{-1}$ are plotted in Fig.~\ref{fig:slepton-constraints} as black lines, where we show both the pure Drell-Yan contribution (means $\lambda_{HS}=0$, dashed) and the inclusion of gluon-gluon fusion results for $\lambda_{HS}=3$ (dotted). For large $\lambda_{HS}$, the reach can be visibly enhanced. Note that the enhancement is not significant in the low $m_S$ region, although in that region the gluon-gluon fusion process has a larger fraction, as shown in Fig.~\ref{fig:SS_LHC}. That is because the $S^\pm$ from gluon-gluon fusion typically have a softer $p_T$, and hence they are cut away by the hard $m_{T2}$ cut in our simulation. Loosing $m_{T2}$ might help to probe the low $m_S$ region, and we leave the detailed study for a future work.

\subsubsection{$e^+e^- \to S^{\pm} S^{\mp(*)}$ at future $e^+e^-$colliders}

As shown in Fig.~\ref{fig:slepton-constraints}, there is a gap between the LHC and LEP constraints for $100~{\rm GeV}< m_S \lesssim 150$ GeV and $30~{\rm GeV}\lesssim m_\chi\lesssim100$ GeV. The future $e^+e^-$ colliders with a collision energy of $\sim250$ GeV can fill this gap. Moreover, an $e^+e^-$ machine is able to probe the lepton portal coupling $y_\ell$ directly, provided one $S^\pm$ is off-shell. For the on-shell production at LHC, since $S^{\pm}\to\ell^{\pm}\chi$ decay branching ratio is $100\%$, the rate does not depend on $y_\ell$. Therefore, the exclusion of slepton-like particle $S$ at LHC is shown only in the $m_S$-$m_\chi $ plot, but can not constrain $y_\ell$. However, for the $2\to3$ process $e^+e^-\to S^\pm \ell^\mp\chi$ mediated by an off-shell $S^\mp$, the rate does depend on $y_\ell^2$, opening the window to directly probe the DM portal coupling. 

The analysis is carried out on the typical Higgs factory such as CEPC, ILC or FCC-ee, with $\sqrt{s}=250$ GeV. 
We include the on-shell $2\to 2$ $S^+S^-$ pair production as well as the off-shell $2\to 3$ process.
For $\ell = e$, the $2 \to 2$ process includes both Drell-Yan contribution and the $t$-channel $\chi$
mediated contribution. Therefore, in this special case, the $2 \to 2$ cross section already has
the contribution of $y_e$. When $\ell = \mu, ~ \tau$, the cross section does not depend on $y_\ell$, 
since only Drell-Yan process contributes.
Regarding $2\to 3$ process, there are two advantages to include it. One is probing the $y_\ell$ coupling,
and the other is that we can probe $\sqrt{s}/2< m_S < \sqrt{s}$ region. Note that given $m_S$ and $m_\chi$, $y_\ell$ is determined by the relic abundance of the DM.

\begin{figure}
\centering
\includegraphics[width=0.4 \columnwidth]{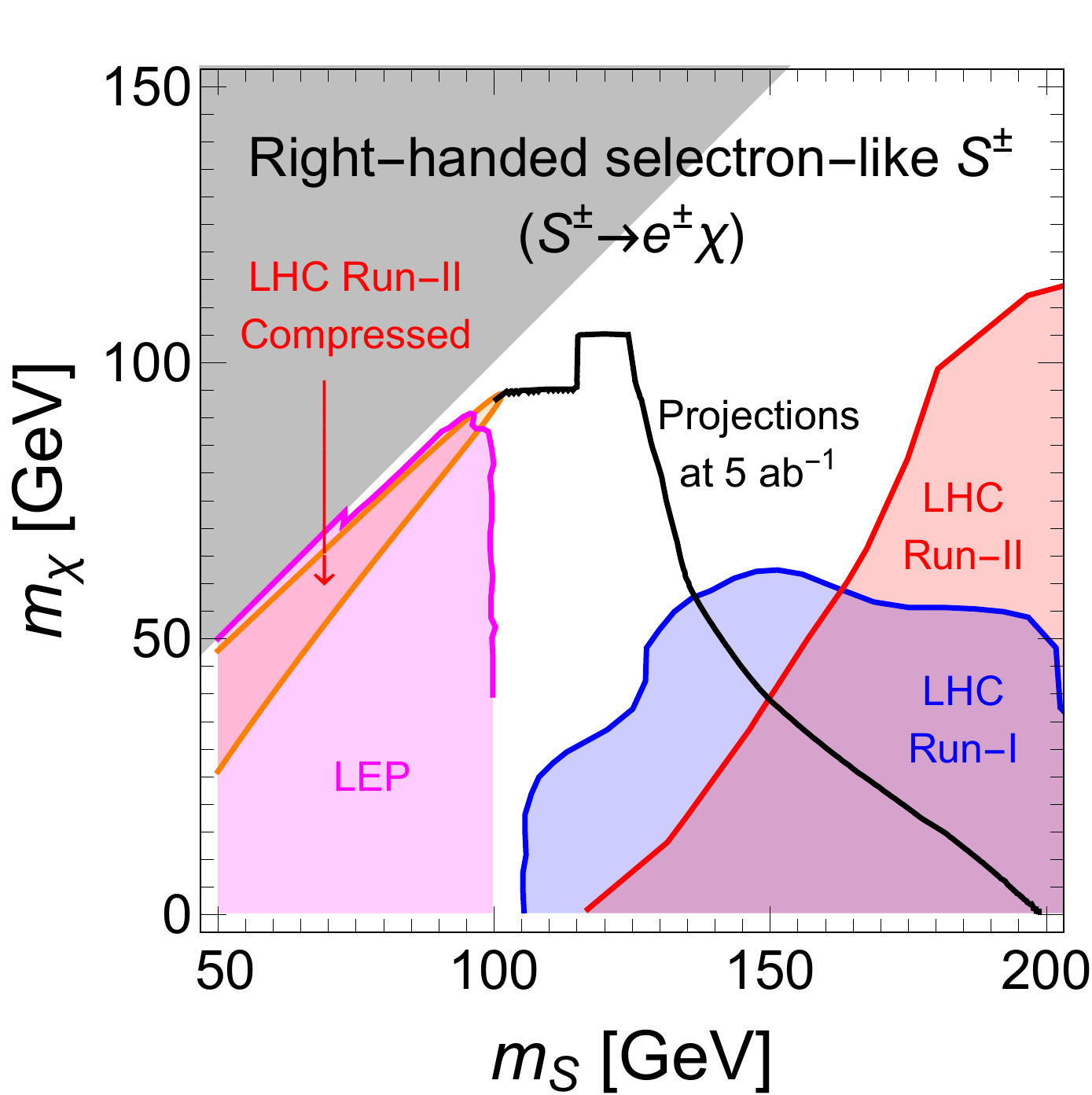}\qquad\qquad
\includegraphics[width=0.4 \columnwidth]{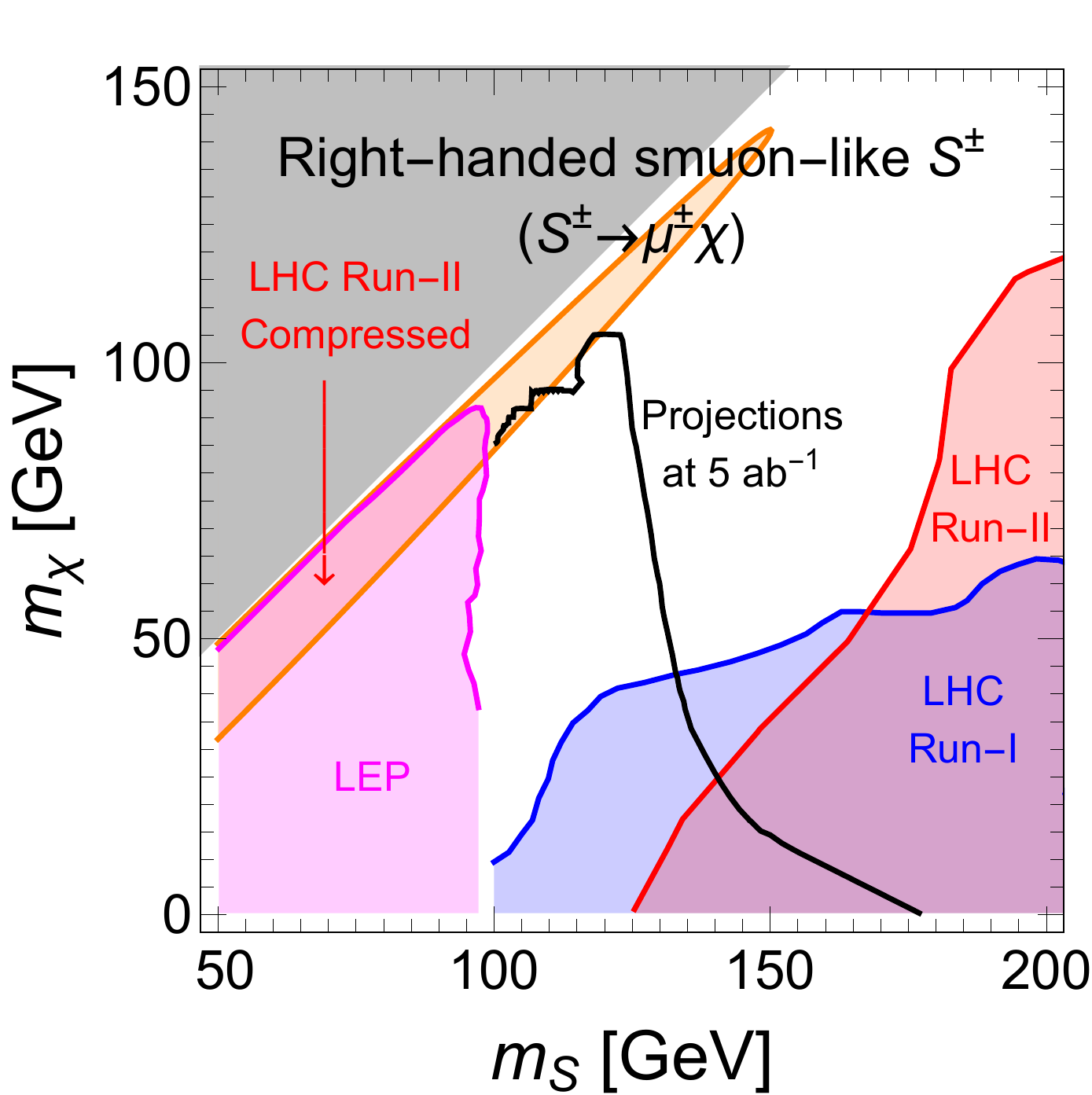}
\caption{The current constraints for right-handed selectron-like and right-handed smuon-like scalar S in the $m_{\rm DM}$-$m_S$ plane. The shaded regions are exclusions from previous experiments, as stated in Fig~\ref{fig:slepton-constraints}. The black lines are projections for the CEPC reach at 5 ${\rm ab}^{-1}$. The zigzag shape of the projections at $m_S\sim120$ GeV are due to the inclusion of off-shell contributions.}\label{fig:CEPC}
\end{figure}

We generate signal events using the packages mentioned in section~\ref{sec:pp_SS}. Following Ref.~\cite{Cao:2018ywk}, we add a few cuts to select the signal events, i.e.
\begin{enumerate}
\item Exactly two opposite charged leptons with $p_T>5$ GeV and $|\eta|<3$;
\item Veto any events with a jet within $p_T>5$ GeV and $|\eta|<3$;
\item The transverse missing momentum $\met>5$ GeV;
\item $m_{T2}>20$ GeV;
\item The polar angles of the leptons satisfy $\cos\theta_{\ell^+}<0.3$ and $\cos\theta_{\ell^-}>-0.3$.
\end{enumerate}
The background event numbers are adopted from the simulated results from Ref.~\cite{Cao:2018ywk}. In Fig.~\ref{fig:CEPC}, we can see that this search is complementary with the LEP and LHC results. It can cover the region $100~{\rm GeV}< m_S \lesssim 150$ GeV region, which is not touched previously when $m_\chi$ mass is moderately large, e.g. $30\sim100$ GeV. For large $m_\chi$, the visible energy shared in the leptons decreases, which makes it hard to compete the LHC background from $W^+ W^-$ pair. The cleaner environment at the lepton collider makes it sensitive to softer leptons comparing with LHC. Including $2\to 3$ process, we do see the sensitivity at future $e^+ e^-$ collider extends to $m_S \sim 170$ GeV and $150$ GeV
for $y_e$ and $y_\mu$ respectively. For small $m_\chi$ there is higher reach for $m_S$, because in this region a large $y_\ell$ is needed to get the correct DM abundance (see \Eq{eq:MajDM}), which enhances the signal significance at the collider (through the off-shell $S^\pm$).

\subsection{Exotic decays from the Higgs and $Z$ bosons}

\subsubsection{Exotic decay: $h/Z\to S^{\pm (*)} S^{\mp (*)}\to\ell^+\chi\ell'^-\chi$}

A charged $S^\pm$ with $m_S < 100$ GeV is already excluded by the LEP experiment, forbidding the $h\to S^+S^-$ one-shell decay. However, for $m_\chi<m_h/2\approx62.5$ GeV, the $\lambda_{HS}$ portal coupling can induce the exotic three- or four-body decays $h \to S^{\pm}\ell^\mp\chi$ or $h\to\ell^\pm\chi\ell^\mp\chi$ mediated by one or two off-shell $S^{\pm}$, depending on whether $m_S<m_h$ or not. The decay width is proportional to $y_\ell^2\lambda_{HS}^2$ or $y_\ell^4\lambda_{HS}^2$, providing a new way to probe $\lambda_{HS}$ and $y_\ell$. If we fix $y_\ell = y_\ell^{\rm th}$ by the relic abundance requirement, it gives a limit to $\lambda_{HS}$ for a given set of $(m_S,m_\chi)$. 

We explore this exotic Higgs decay at the Higgs factory FCC-ee and CEPC at $\sqrt{s} = 250$ GeV via the $e^+ e^- \to Z h$ production channel, whose cross section is 0.24 pb. We consider the following cascade decays $Z \to \ell''^{+} \ell''^{-}$ and $h \to S^{\pm(*)} S^{\mp(*)}\to\ell^+\chi\ell'^-\chi$,\footnote{Both $\ell$ and $\ell'$ denote the lepton $e$ and $\mu$.} where $\chi$ plays the role of missing energy. The main SM backgrounds include $ZW^+W^-$ and $Z\tau^+\tau^-$, with all the particles decaying to leptonic final states, and the $Zh$, with $h$ decaying to $W^\pm W^{\mp*}$, $ZZ^*$ and $\tau^+\tau^-$. The total cross section for the backgrounds in the  $(\ell''^{+} \ell''^{-}) \ell^+ \ell'^- \slashed{E}_T$ final state is as small as 0.02 fb, where the leptons in the parentheses come from $Z$ decay. The signal and background events are simulated by the packages mentioned in section~\ref{sec:pp_SS}. We apply the following detailed requirements to the events,
\begin{enumerate}
	\item At least four charged leptons with $p_T>10$ GeV and $|\eta|<2.47$;
	\item A pair of leptons with same-flavor and opposite-sign, and satisfies $|m_{\ell''^+\ell''^-}-m_Z| < 5~{\rm GeV}$;
	\item The missing energy $\slashed{E}_T>20$ GeV;
	\item The reconstructed Higgs resonance in the mass window~\cite{Liu:2016zki}
	\be
	120~{\rm GeV}< \sqrt{\left(p_{e^+}+p_{e^-}-p_{\ell''^+}-p_{\ell''^-}\right)^2} < 130~{\rm GeV},
	\ee
	which is equivalent to cut on the total energy of $Z$.
\end{enumerate}

\begin{figure}
\centering	
\includegraphics[width=0.4 \columnwidth]{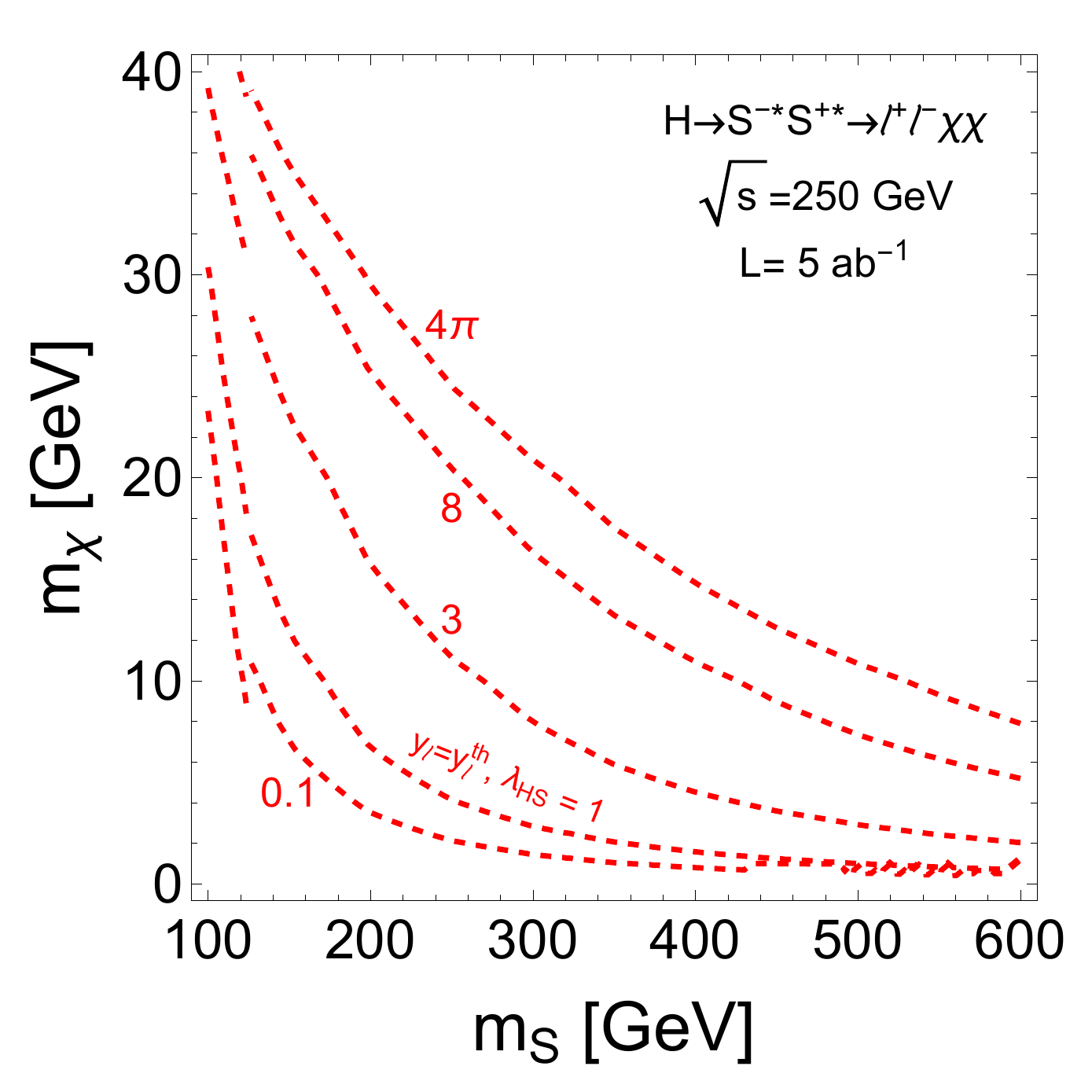}\qquad\qquad
\includegraphics[width=0.4 \columnwidth]{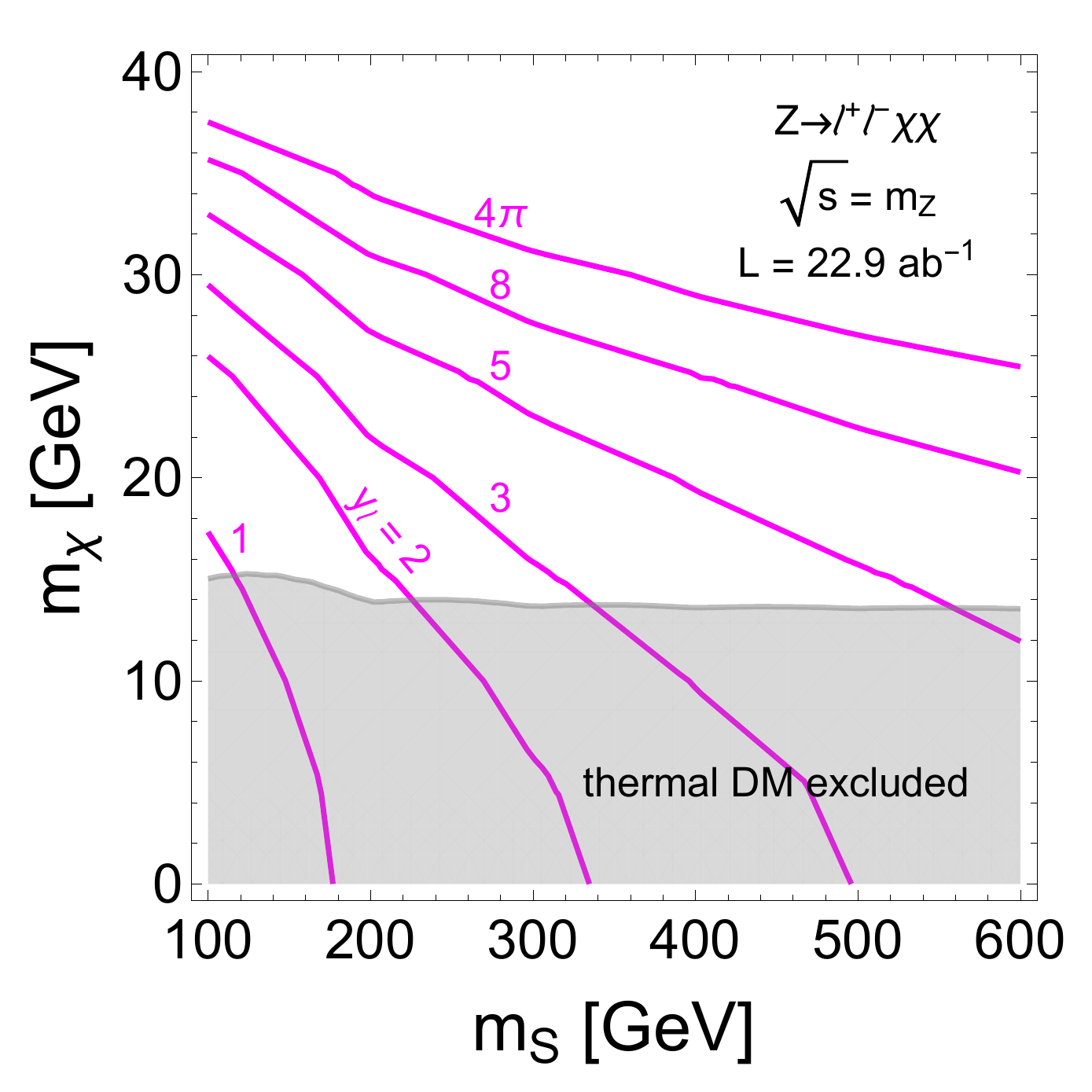}
\caption{Left: The $95\%$ C.L. constraint on the coupling combination $\lambda_{HS}$ as contours from the exotic decay $h \to S^{\pm (*)} S^{\mp*}\to\ell^+\chi\ell'^-\chi$. The lepton portal coupling $y_\ell$ is set to be $y_\ell^{\rm th}$, which is the value to provide the correct DM relic abundance. Right: The $95\%$ C.L. constraint contours (magenta) for $y_\ell$ for the exotic decay $Z\to\ell^+\chi\ell'^-\chi$. The gray shaded region shows the parameter space can be excluded if $y_\ell=y_\ell^{\rm th}$.}\label{fig:ycoupling-from-hZ}
\end{figure}

After the cuts, for a given integrated luminosity we are able to set bounds for $\Br(h\to S^{\pm(*)} S^{\mp(*)}\to\ell^+\chi\ell'^-\chi)$, which in turn is translated into the upper limits for $\lambda_{HS}$ once $y_\ell$ is fixed by the relic abundance requirement, as shown in left panel of Fig.~\ref{fig:ycoupling-from-hZ}. The discontinuity of the curves around $m_S = 125$ GeV is originated from the phase space change from three-body to four-body decay. In conclusion, the future $e^+ e^-$ collider can significantly constrain the couplings for light $m_\chi \lesssim 30$ GeV and $m_S$ of a few hundreds GeV.

In addition to the exotic SM Higgs decay, another good target to probe new physics is the $Z$ exotic decay \cite{Liu:2017zdh}. To explore the SM parameters with better precision, the future $e^+ e^-$ colliders have the proposals to run at $Z$-pole \cite{CEPCStudyGroup:2018ghi, Abada:2019lih, Abada:2019zxq}, which can provide Giga ($10^9$) or Tera ($10^{12}$) $Z$ boson. Interestingly, in this model, there exists the exotic decay channel $Z\to\ell^+\chi\ell'^-\chi$, which provides a di-lepton plus missing energy final state. There are two types of diagrams responsible for this channel, the first one involves two off-shell $S^{\pm}$ through $ZS^+S^-$ coupling and the second one involves one off-shell $S^{\pm}$ through $Z\ell^+\ell^-$ coupling with $S$ attached to one of the charged lepton.
Since we are considering $S^\pm$ with mass larger than 100 GeV, it is heavier than all other particles. Therefore,
the decay width is dominated by the latter diagram, which is proportional to $y_\ell^4 m_S^{-4}$. 

We explore this exotic $Z$ decay similar to the Higgs case. The dominant SM background is $\bar{\nu}\nu \ell^+ \ell'^-$ from off-shell gauge boson pair production.\footnote{The background from tau pair provides softer leptons due to more neutrinos in the final states, and can be further suppressed by requiring large $m_{T2}$.} The cut conditions are:
\begin{enumerate}
\item At least two same-flavor opposite-sign leptons with $p_T> 10$ GeV and $\eta< 2.5$;
\item The missing energy should satisfy $\slashed{E}_T> 10$ GeV.
\end{enumerate}
The $ 95\%$ C.L. constraint  on the exotic decay branching ratio is about $\Br(Z\to e^-e ^+ \chi\chi) \lesssim 10^{-9}$ for Tera $Z$ option, while exact limit again depends on $m_\chi$ and $m_S$. In right panel of Fig.~\ref{fig:ycoupling-from-hZ}, we show the $ 95\%$ C.L. constraint on $y_\ell$ presented as the number above the magenta contours. Moreover, we compare this constraint with the thermal relic requirement $y_\ell^{\rm th}$. It shows that for Tera $Z$ option, the thermal DM with mass $m_\chi \lesssim 13 $ GeV can be excluded by this exotic $Z$ search, 
because $y_\ell^{\rm th}$ exceeds the limit from the exotic $Z$ decay $Z\to e^-e ^+ \chi\chi$. 
We plot such region in gray and label as ``thermal DM excluded".
This constraint provides a complimentary limit for large $m_S$ comparing with LHC limits, because it does not require on-shell $S$ production. Moreover, since the decay width and the DM annihilation cross section are proportional to  
$y_\ell^4 m_S^{-4}$, this exclusion line can extend horizontally to very high $m_S$, thus is a powerful constraint
for this DM model.

\subsubsection{Invisible decay: $h\to\chi\chi$}

\begin{figure}
\centering	
\includegraphics[width=0.32 \columnwidth]{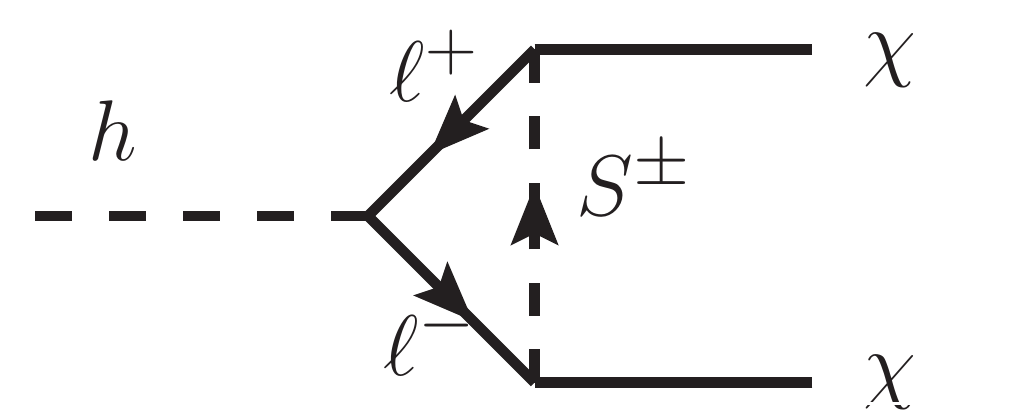}\qquad
   \includegraphics[width=0.32 \columnwidth]{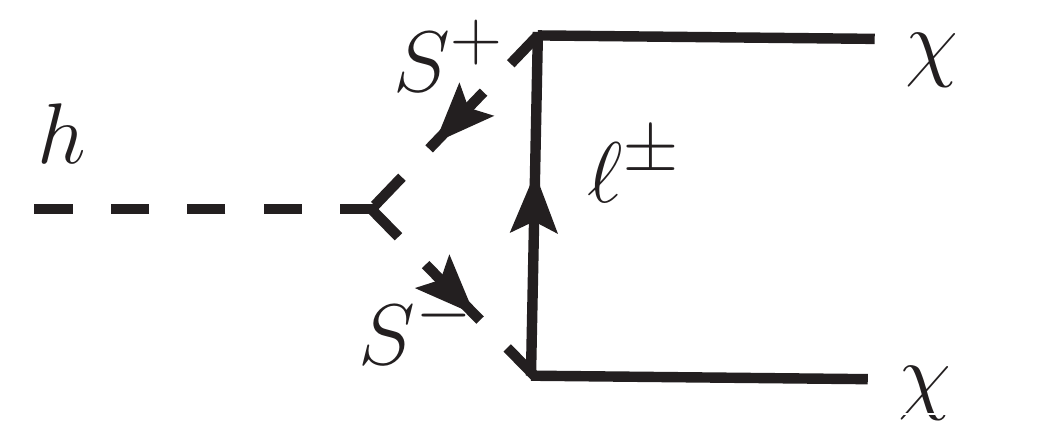}
\caption{The one-loop induced Higgs invisible decay. The cross-diagrams for Majorana fermion $\chi$ are not shown here, but are included in the calculation.}\label{fig:h_chichi}
\end{figure}

The Higgs invisible decay $h \to \chi \chi$ is induced at one-loop level by the two Feynman diagrams listed in Fig.~\ref{fig:h_chichi}.\footnote{This is similar to the Higgs to neutralinos decay in the SUSY models \cite{Drees:1996pk, Djouadi:2001kba, Eberl:2001vb, Berlin:2015njh}.} The first diagram is negligible due to the small lepton mass. We calculate the second diagram contribution to the exotic Higgs decay using {\tt Package-X}~\cite{Patel:2015tea}. The induced coupling is
\be
\mathcal{L}_{\rm 1-loop}^\chi \supset  \frac{1}{2} g_{h\chi \chi} h \bar{\chi} \chi,
\ee
where
\begin{align}
 &	g_{h\chi \chi}
	= - \frac{y_\ell^2\lambda_{HS} m_\chi v}{4 \pi^2 (4 m_\chi^2 - m_h^2)} \times  \label{eq:1loop-hchichi} \nn\\
	&\left[ \text{DiscB}(m_h^2,m_S,m_S) +\frac{m_S^2-m_\chi^2}{m_\chi^2}\left(\log\frac{m_S^2}{m_S^2-m_\chi^2}-m_\chi^2 C_0(m_h^2,m_\chi^2,m_\chi^2,m_S,m_S,0) \right)\right] , \nonumber \\
	& \approx  - \frac{ y_\ell^2 \lambda_{HS}m_\chi v}{16 \pi^2  m_S^2 }  
	+\mathcal{O}\left(m_S^{-3} \right)
\end{align}
where $\text{DiscB}$ is the finite part of the Passarino-Veltman function $B_0$ defined in {\tt Package-X}, $C_0$ is the Passarino-Veltman function following the definition in {\tt Package-X} and the lepton mass is taken to be zero. In the second equality, we have expanded the result with large $m_S$. We have checked our result with Ref.~\cite{Djouadi:2001kba} and Ref.~\cite{Berlin:2015njh}, and found agreement between each other. 
In numeric calculation, we use the full expression from Eq.~(\ref{eq:1loop-hchichi}).

The $h\to \chi\chi$ partial width is given as
\begin{align}
	\Gamma(h\to \chi\bar\chi)=\frac{g_{h\chi\chi}^2m_h}{8\pi}\left( 1-\frac{4m_\chi^2}{m_h^2}\right)^{\frac{3}{2}} \,.
\end{align}
The current best limit for invisible Higgs decay is $\Br(h \to {\rm inv}) < 13\%$  at ATLAS Run-II with integrated luminosity $139~{\rm fb}^{-1}$ \cite{ATLAS-CONF-2020-008}. For future HL-LHC, the sensitivity for invisible Higgs BR is $3.5\%$ from \cite{Bernaciak:2014pna}. For future $e^+e^-$ collider such as CEPC, the sensitivity can be increased to about $0.3\%$ \cite{CEPCStudyGroup:2018ghi}.

\begin{figure}
\centering	
\includegraphics[width=0.4 \columnwidth]{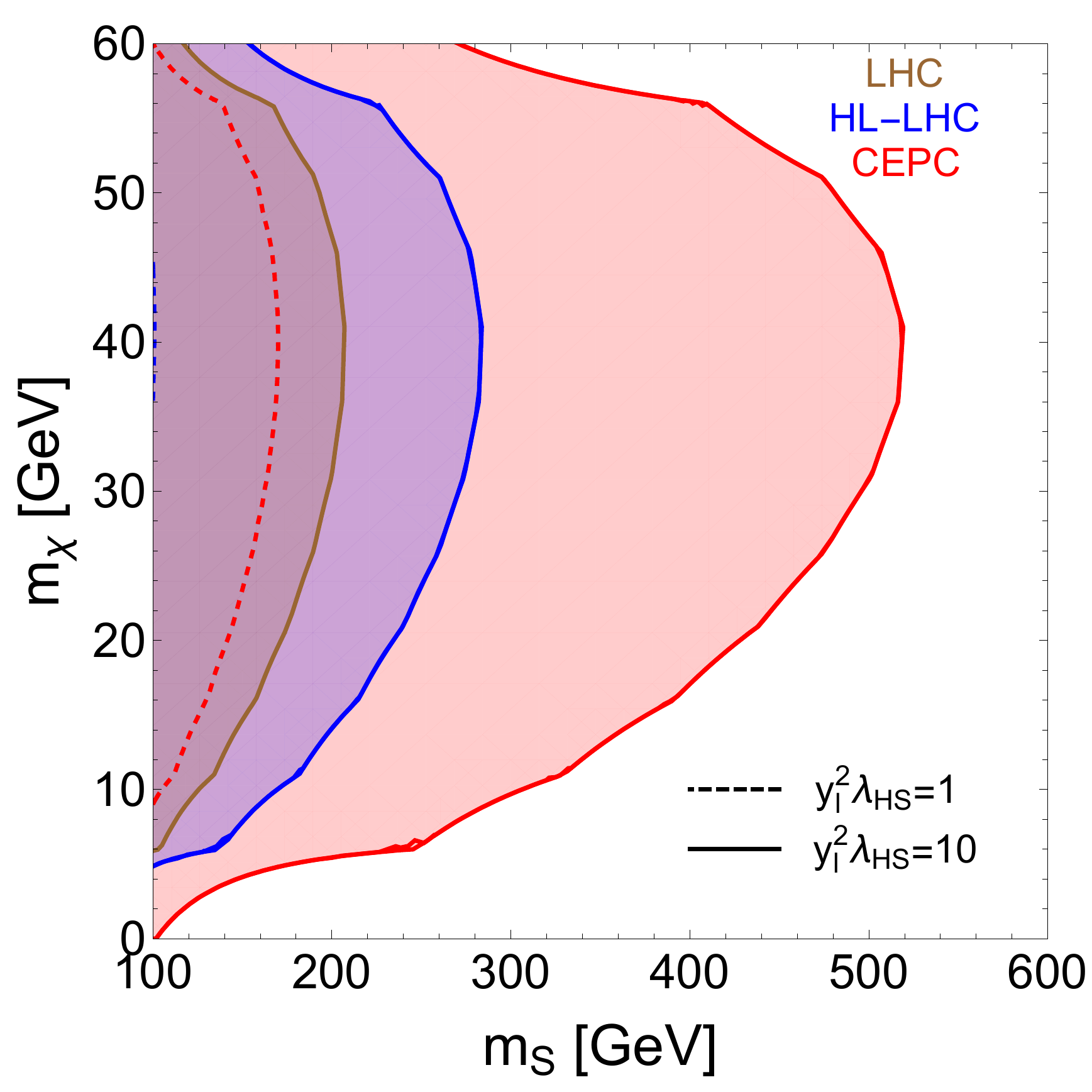}\qquad\quad
\includegraphics[width=0.45 \columnwidth]{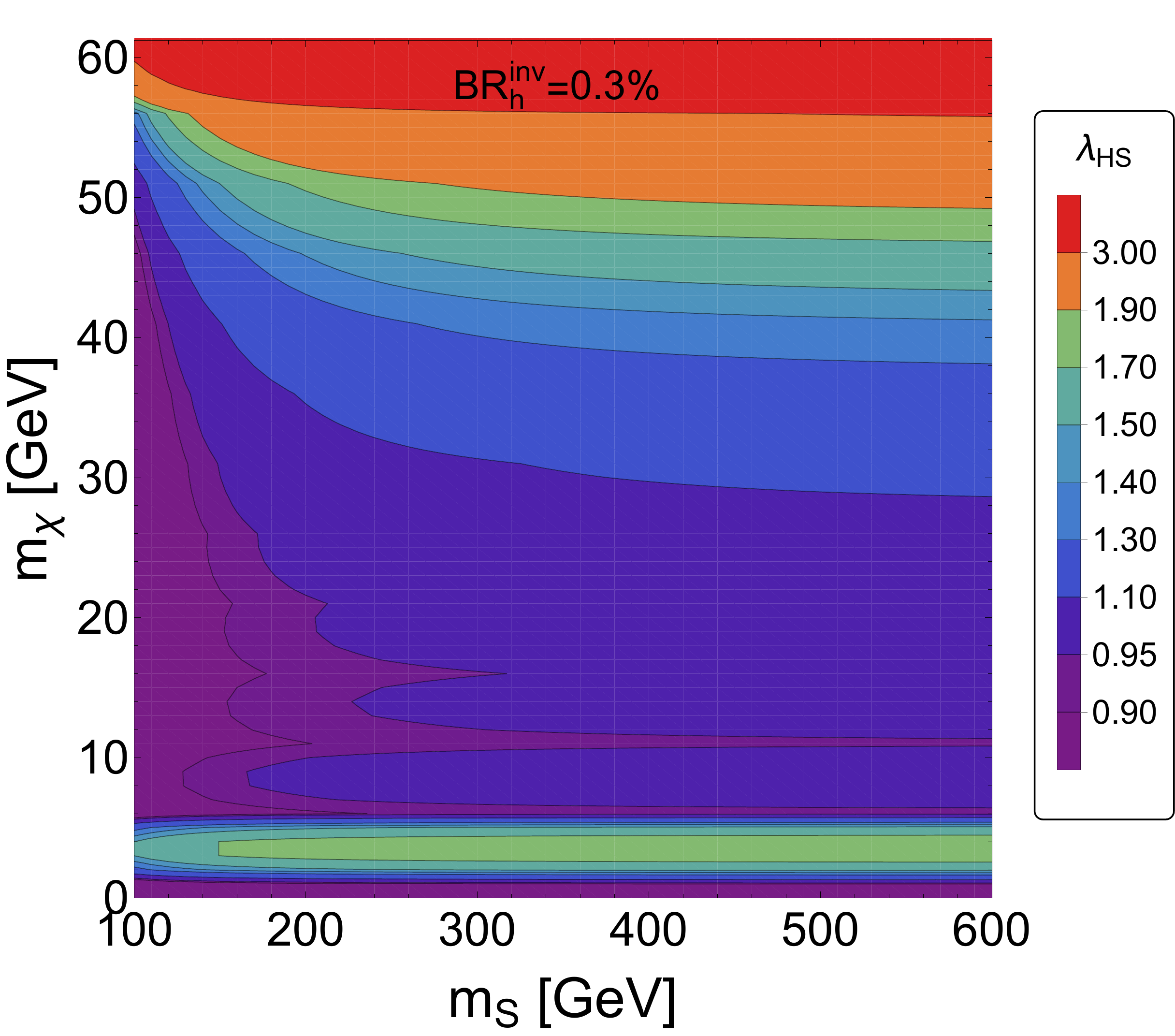}
\caption{Left: the Higgs invisible decay as a probe for the combination $y_\ell^2\lambda_{HS}$ at different colliders. Right: the CEPC sensitivity $\Br( h \to {\rm inv}) < 0.3\%$ can set bounds on $\lambda_{HS}$ if we fix $y_\ell = y_\ell^{\rm th}$ from the relic abundance requirement.}
\label{fig:lambdaHS-from-BRinv-CEPC}
\end{figure}

The above projections from future colliders can set limits on the coupling combination $y_\ell^2\lambda_{HS}$ as a function of masses $m_S$ and $m_\chi$. In the left panel of Fig.~\ref{fig:lambdaHS-from-BRinv-CEPC}, we show the contours of $y_\ell^2\lambda_{HS}$ for LHC (brown), HL-LHC (blue) and CEPC (red) sensitivities. The dashed and solid contours corresponds to $y_\ell^2\lambda_{HS}= 1,~10$ respectively.
We clearly see that the future $e^+ e^-$ collider has a better sensitivity over the hadron colliders.
In the right panel of Fig.~\ref{fig:lambdaHS-from-BRinv-CEPC}, $y_\ell$ is set to be $y_\ell^{\rm th}$ to satisfy 
DM relic abundance requirement. Once fixing $m_S$ and $m_\chi$, the future sensitivity on $\lambda_{HS}$
can be calculated using CEPC sensitivity $\Br(h\to{\rm inv}) = 0.3\%$ and the contours are shown.
One interesting feature is that for $m_\chi$ smaller than 6 GeV, the sensitivity on $\lambda_{HS}$ goes down significantly, because
the leading order in the width for small $m_\chi$ expansion is linear in $m_\chi$. When $m_\chi$ decreases further, $y_\ell^{\rm th}$
increases to compensate the annihilation cross section, which makes the sensitivity on $\lambda_{HS}$ becoming stronger again. Therefore, we see $\lambda_{HS}$ reaches its best sensitivity for small $m_S$ and moderate $m_\chi$.

\subsection{One-loop contributions to Higgs couplings}
\label{sec:1loop}

\begin{figure}
\centering
\includegraphics[width=0.32 \columnwidth]{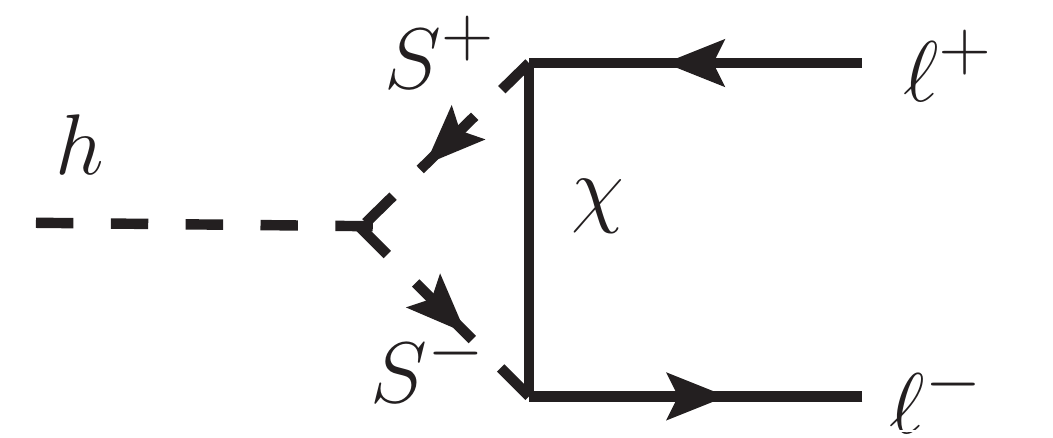}
\caption{The one-loop induced $h\ell^+\ell^-$ coupling modification.}
\label{fig:h_ll}
\end{figure}

The $h\ell^+\ell^-$ vertex is modified by the one-loop diagram in Fig.~\ref{fig:h_ll},\footnote{Similar corrections have been
studied in SUSY model already~\cite{Djouadi:2001kba, Heinemeyer:2015pfa}.} and the induced interaction is calculated using the {\tt Package-X}
\begin{align}
& \mathcal{L}_{\rm 1-loop}^{\ell}  \supset  
-\frac{ y_{\ell}^2\lambda_{HS} v m_{\ell}}{8\pi^2 m_h^2}h \bar{\ell} \ell   
\label{eq:1loop-Hll} \\
&\times \left[ 1+\text{DiscB}(m_h^2,m_S,m_S)+ \frac{m_\chi^2}{m_S^2-m_\chi^2}\log\frac{m_S^2}{m_\chi^2}-\left( m_S^2-m_\chi^2\right)C_0(0,0,m_h^2,m_S,m_\chi,m_S)\right], \nonumber
\end{align}
where we have taken the leading power in small lepton mass $m_\ell$.
It is obvious that the coupling is lepton mass suppressed, because the DM only couples to right-handed lepton and one has to flip the helicity of lepton, which induce this suppression. Of course, the SM Higgs couplings to leptons are suppressed by the mass as well. Thus, the one-loop correction to the SM coupling in fraction is proportional to $y_\ell^2\lambda_{HS}/(8 \pi^2)$. In addition to vertex correction, the Higgs field renormalization and the lepton field renormalization can also contribute to the coupling. The experiments characterize the sensitivity of Higgs coupling measurements in the $\kappa$-framework~\cite{Cepeda:2019klc}, which is calculated using the cross section $\sigma(Zh)$ and $H$ decay branching ratios and hence the universal Higgs field renormalization effect cancels out in the decay branching ratios. Therefore, Higgs field renormalization does not affect the branching ratio to leptons. For the lepton field renormalization, it only contributes to right-handed lepton. We calculated that it is about $y_\ell^2/(128 \pi^2)$ for large $m_S$ expansion, which is much smaller than Eq.~\eqref{eq:1loop-Hll}. Together with the fact that $\lambda_{HS} \gtrsim 1$, we can neglect the right-handed lepton field renormalization effect and focus on the vertex correction.

The most recent constraints from LHC for Higgs decay to leptons are $\Br(h\to\mu^+ \mu^-)< 3.8\times 10^{-4}$~\cite{ATLAS-CONF-2019-028} and coupling ratio $\kappa_\tau =1.05^{+0.16}_{-0.15}$~\cite{Aad:2019mbh}. 
Moreover, we can also consider  the sensitivities at future $e^+ e^-$ collider. For example, 
the relative precision for coupling measurement  from CEPC study in 10-parameter fit are
$\delta \kappa_\mu < 8.7\% $ and $\delta \kappa_\tau < 1.5\% $~\cite{An:2018dwb}, similar to FCC-ee study~\cite{Abada:2019zxq}. Therefore, we convert the above existing constraints and future sensitivities on leptons into the combination of couplings
$y_\ell^2\lambda_{HS}$. 

\begin{figure}
	\centering	
	\includegraphics[width=0.30 \columnwidth]{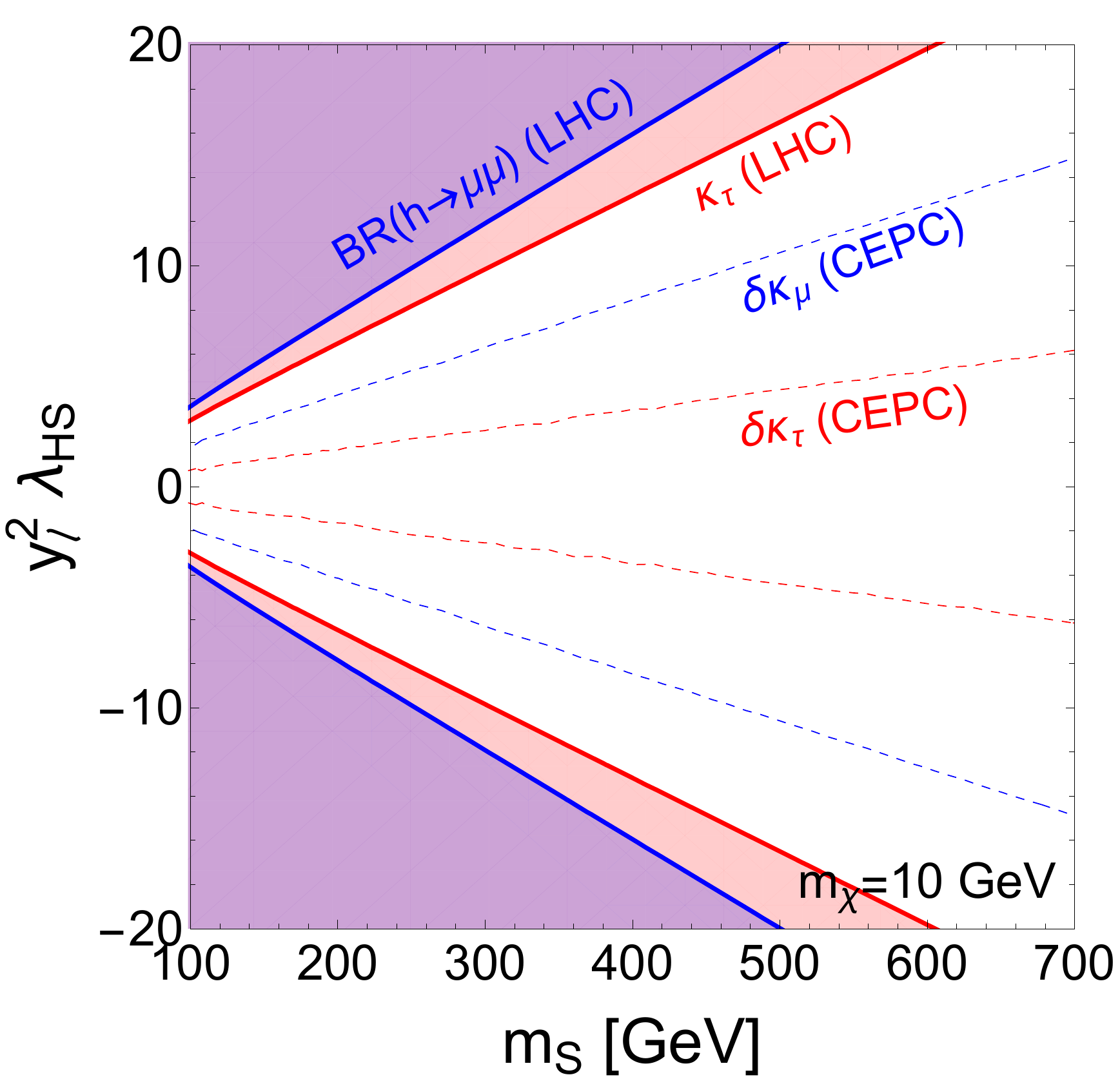}~~
	\includegraphics[width=0.33 \columnwidth]{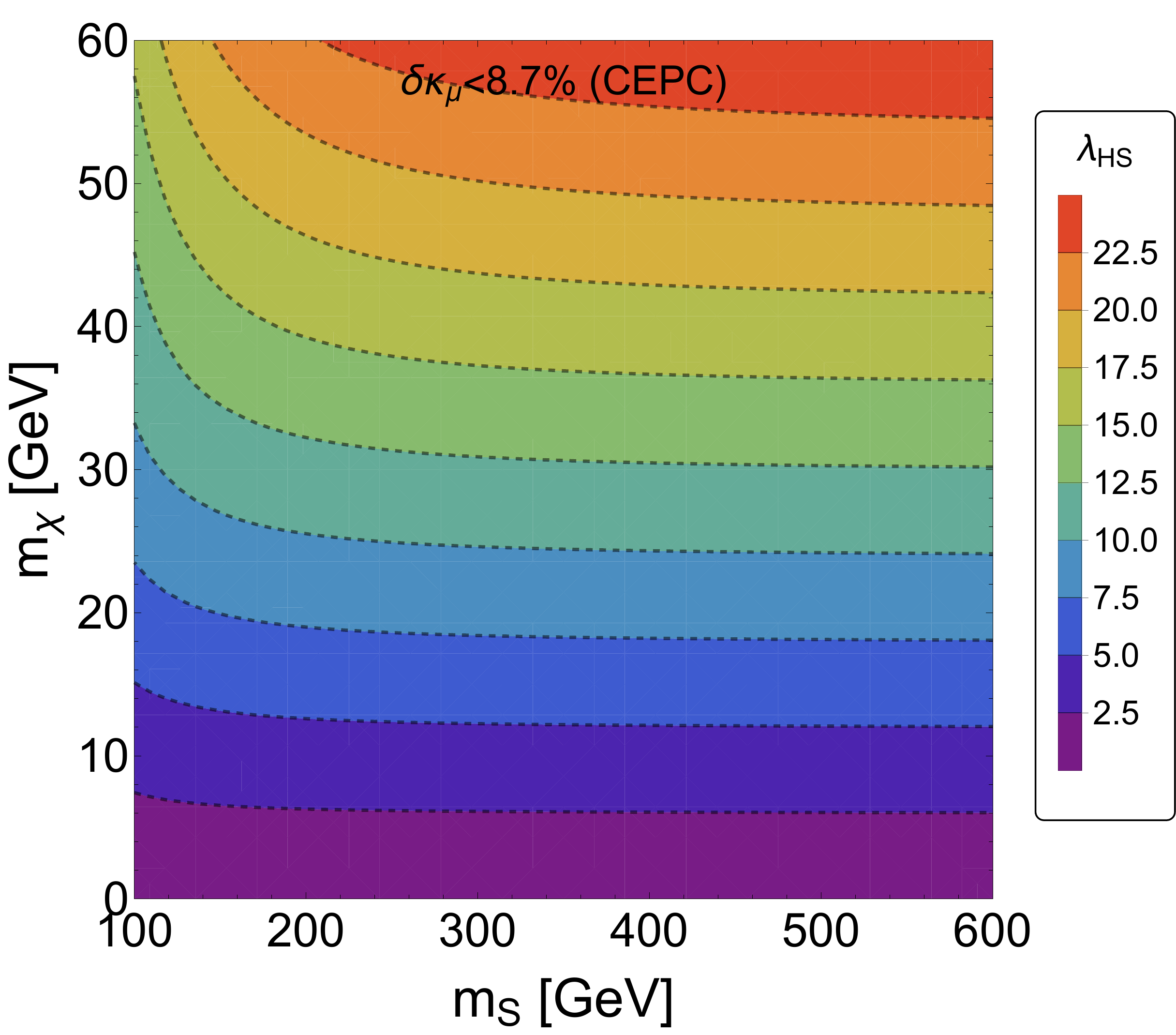}~~
	\includegraphics[width=0.324 \columnwidth]{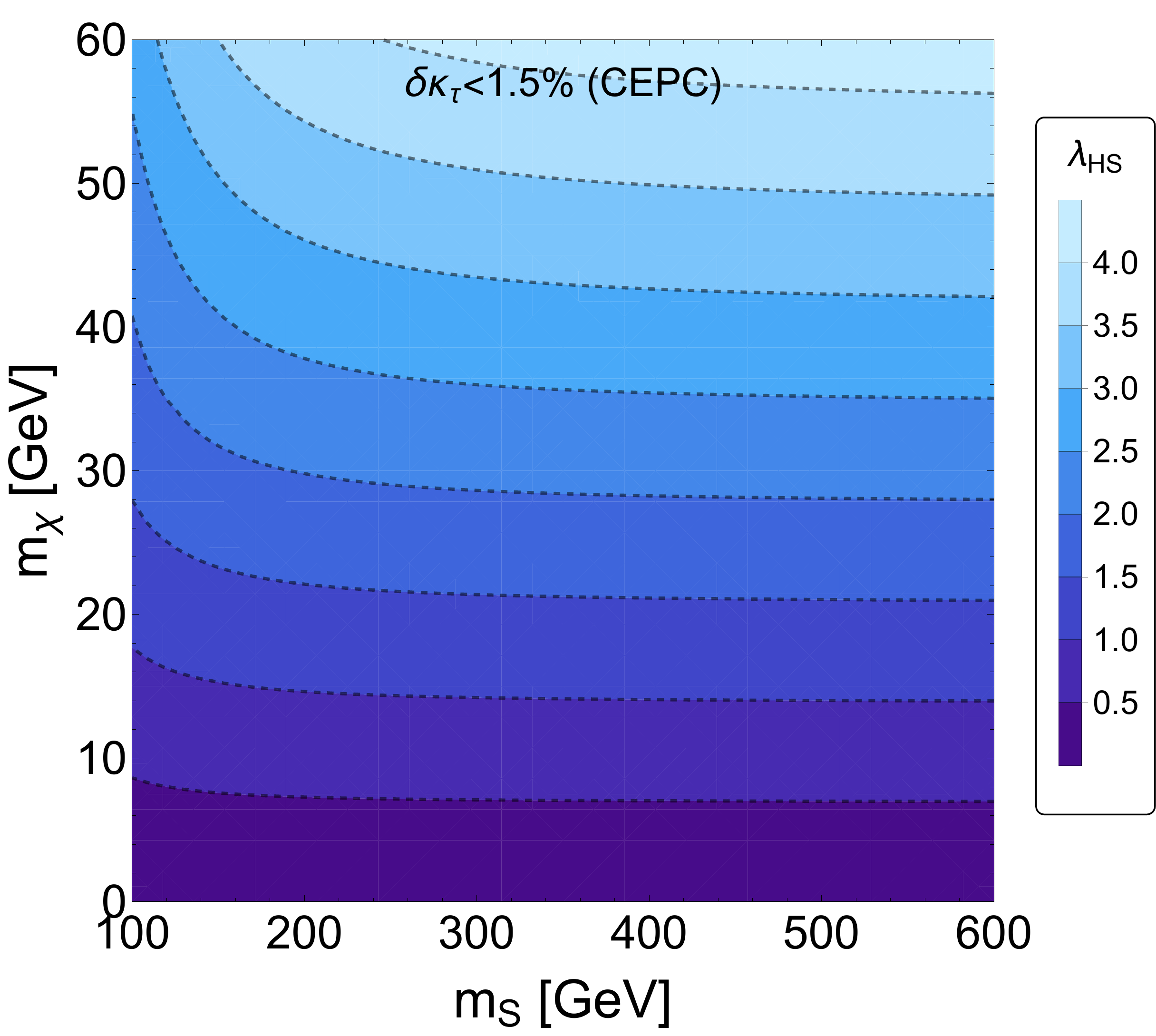}
	\caption{Left panel: The constraints on $\lambda_{HS} y_\ell^2$ from the $h\to \ell\ell$
		branching ratio and coupling strength $\kappa_{\tau}$ from existing LHC constraints \cite{ATLAS-CONF-2019-028, Aad:2019mbh} with fixed DM mass $m_\chi = 10$ GeV. Middle and right panels: the constraint on $\lambda_{HS}$ from the projected precision $\delta \kappa_{\mu} < 8.7\%  $ and $\delta \kappa_{\tau} < 1.5 \%  $ from CEPC~\cite{An:2018dwb}, with $y_\ell = y_\ell^{\rm th}$ by DM relic abundance. }
	\label{fig:lambdaHS-from-H-to-ll}
\end{figure}

The results are shown in Fig.~\ref{fig:lambdaHS-from-H-to-ll}. In the left panel, we show the limits for the combination $y_\ell^2\lambda_{HS}$
from $\Br(h \to \mu^+\mu^-, \tau^+\tau^-$) at LHC and CEPC. The shaded regions are already excluded by current LHC limits, while the future sensitivities from CEPC are plotted as dotted lines.
The DM mass is fixed as $m_\chi = 10$ GeV and the limits get weakened linearly with increasing $m_S$. 
In the middel and right panels, we fix $y_\ell = y_\ell^{\rm th}$ for the DM relic abundance requirement, 
and show the sensitivity contours for $\lambda_{HS}$ as a function of $m_S$ and $m_\chi$.
When taking $m_S \to \infty$, the coupling in \Eq{eq:1loop-Hll} is proportional to $y_\ell^2 m_S^{-2}$. Therefore, 
it has the similar dependence as the DM relic abundance requirement which is $y_\ell^4 m_S^{-4}$. 
This feature is clearly shown that when $m_S$ increases, the contours for $\lambda_{HS}$ are flat. It means that the sensitivity does not suffer for large $m_S$, because the large mass $m_S$ is compensated by large $y_\ell^{\rm th}$.
Since for $\mu$ and $\tau$, this change of coupling ratio $\delta \kappa$ does not depend on lepton mass, the constraint on $\lambda_{HS}$ is only linear depends on the CEPC precision, which has better $\tau$ sensitivity. Therefore, the sensitivity for $\tau$ lepton portal is better by a factor of $5.8$ compared to $\mu$ lepton portal. 

\begin{figure}
\centering	
\includegraphics[width=0.4\columnwidth]{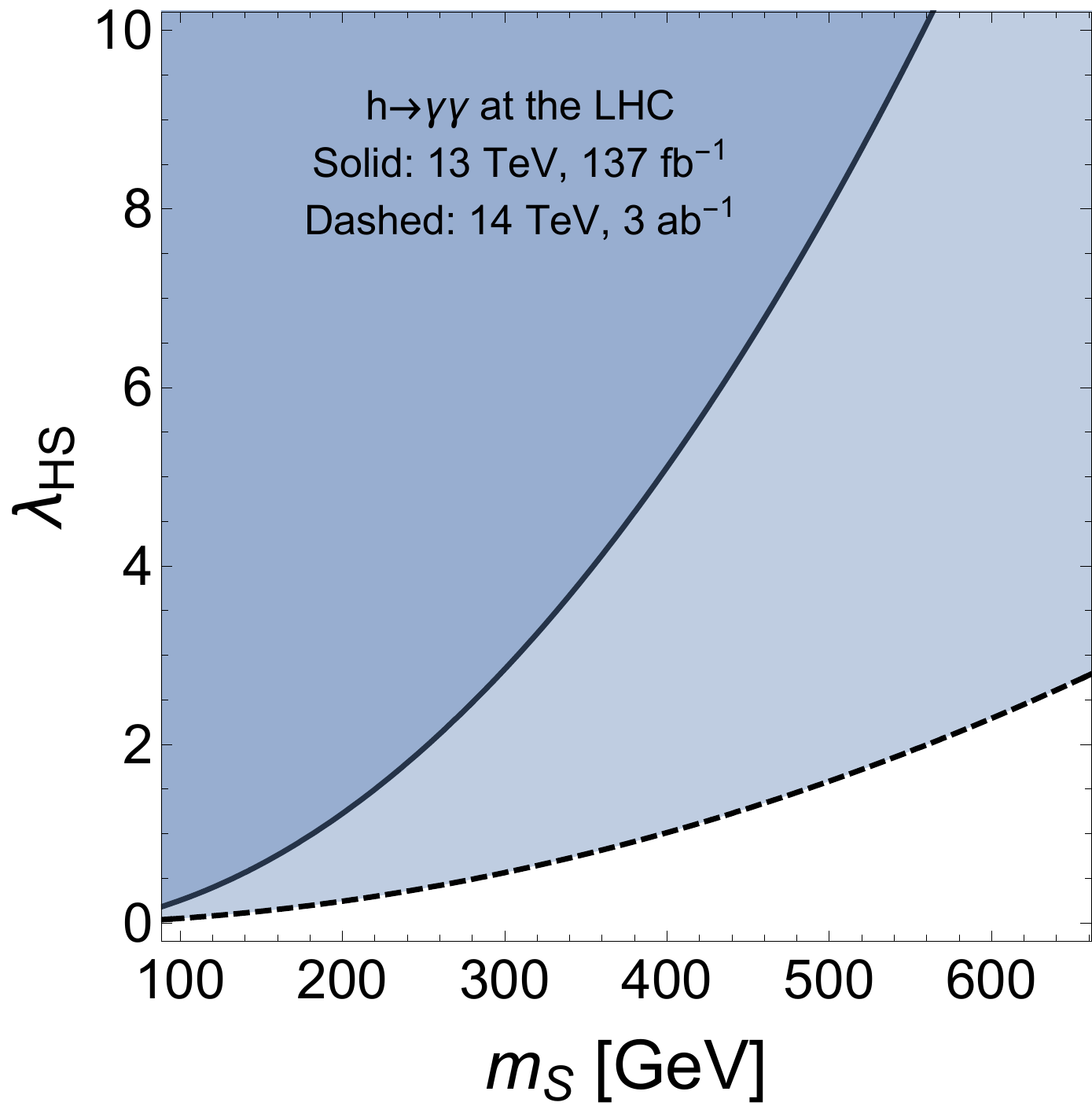}\qquad\qquad
\includegraphics[width=0.4\columnwidth]{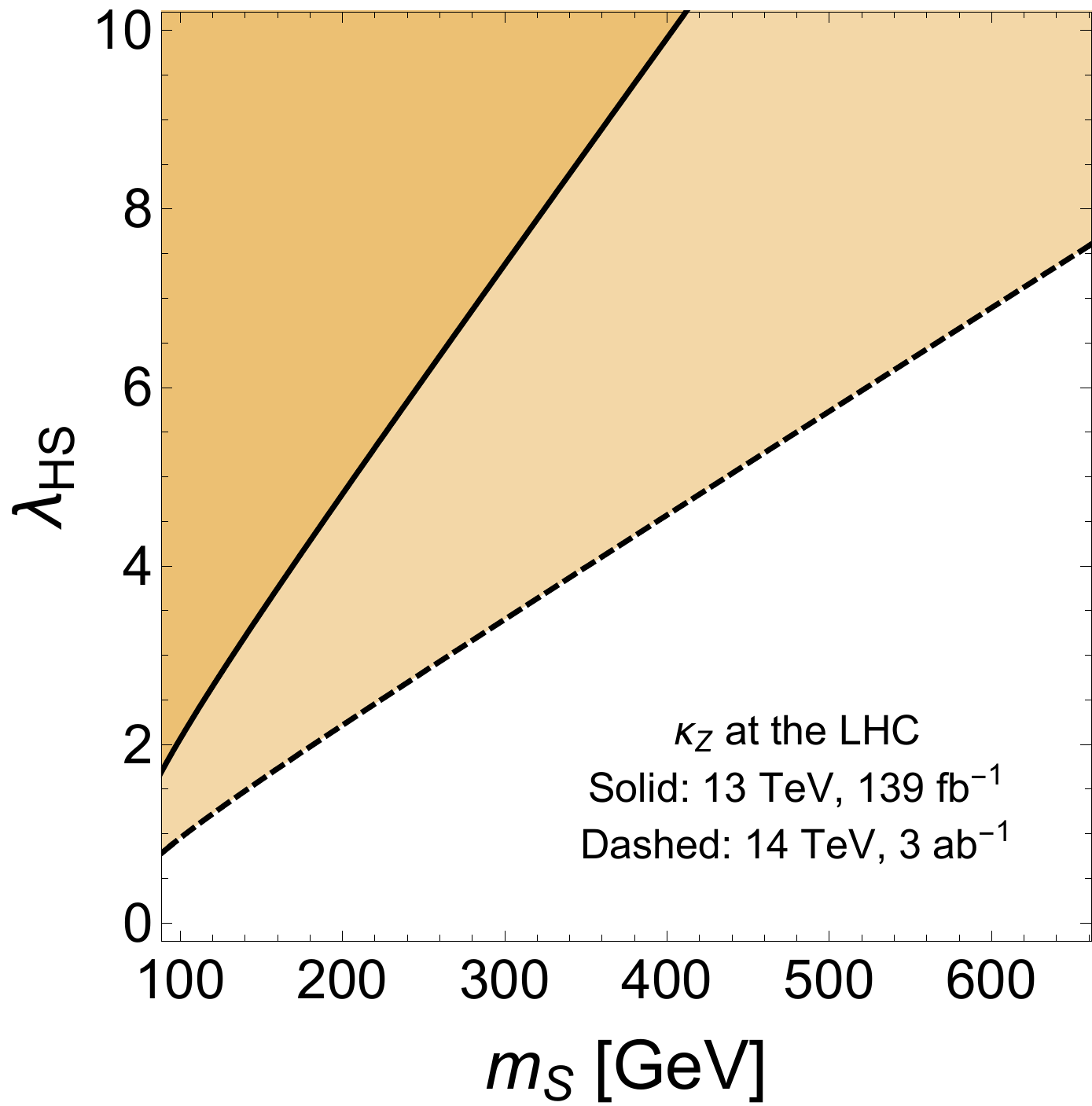}
\caption{The shaded regions are constraints and projections from the $h\gamma\gamma$ and $hZZ$ coupling measurements at the LHC.}
\label{fig:haa_hZZ}
\end{figure}

Besides the one-loop modification to $h\ell^+\ell^-$ coupling, the charged scalar loop can also modify the $h\gamma\gamma$ and the $hZZ$ couplings~\cite{Huang:2016cjm}. The former one modifies $\Br(h \to \gamma \gamma)$ via the charged scalar triangle loop, 
and current fitting result for the signal strength is $1.12\pm0.09$ at the 13 TeV LHC with 137 fb$^{-1}$~\cite{1851456}, while the projection accuracy is 4\% at the HL-LHC~\cite{Cepeda:2019klc}, better than the projected sensitivity 6.9\%
at the 5.6 ab$^{-1}$ CEPC~\cite{CEPCStudyGroup:2018ghi}.
The latter one is mostly contributed by Higgs field renormalization (via the $S^\pm$ loop) if $\lambda_{HS} > e$ (the elementary charge). This modification doesn't rely only on $y_\ell$, therefore it provides a direct constraint on the scalar interaction coupling $\lambda_{HS}$ independent of the Yukawa coupling. Current constraint for $\delta\kappa_Z$ is 8\% for the 13 TeV 139 fb$^{-1}$ LHC~\cite{Aad:2020mkp}, and the projected result is 1.7\% at the HL-LHC~\cite{Cepeda:2019klc}.
Therefore, the precision measurement on $Zh$ production cross section can provide a limits on $\lambda_{HS}$. Following the procedure of Ref.~\cite{Huang:2016cjm}, we calculate the sensitivity for $\lambda_{HS}$ from the future limits on $\Br(h \to \gamma \gamma$)
and the coupling strength to $Z$ gauge boson $\kappa_Z$. The results are presented in Fig.~\ref{fig:haa_hZZ}.
 
\subsection{The lepton $g-2$}

Another consequence of the lepton portal coupling $y_\ell$ is the magnetic dipole moment for leptons at one-loop level. The $g-2$ contribution for lepton is given as \cite{Moroi:1995yh, Carena:1996qa}
 \begin{align}
 \Delta a_{\ell} \equiv a_\ell^{\rm exp} - a_\ell^{\rm SM}  = - \frac{y_\ell^2}{16 \pi^2} \frac{m_\ell^2}{m_S^2} \frac{1- 6 x + 3 x^2 + 2 x^3 - 6x^2 \log x}{6 (1-x)^4},
 \label{eq:g-2}
 \end{align}
where $x  \equiv m_\chi^2/m_S^2$ and we keep the leading order result in the limit of small $m_\ell$.
The last term containing $x$ in Eq.~(\ref{eq:g-2}) is a monotonically decreasing function of $x$ and it equals to $1/6$ and $1/12$ in the limit of $x\to 0$ and $x\to 1$. Moreover, $ \Delta a_{\ell}$ is always negative. 
The electron magnetic dipole moment has been directly measured in Refs.~\cite{Hanneke:2008tm, Hanneke:2010au}, and 
one can compare it with the SM prediction~\cite{Aoyama:2014sxa, Mohr:2015ccw} once the fine structure constant 
is given. The recent results are
\be
\Delta a_e = (-88 \pm 36)\times 10^{-14} , \quad \Delta \tilde{a}_{e}=(48 \pm 30)\times 10^{-14},
\ee
where $\Delta a_e $ comes from the $\alpha$ measurement in recent cesium recoil experiments~\cite{Parker:2018vye},
while $\Delta \tilde{a}_e $ comes from a new independent measurement using rubidium atom~\cite{Morel:2020dww}.
For electron $g-2$, the uncertainties from QED, EW and hadronic contributions are much smaller than $\alpha$,
while for muon $g-2$, the uncertainty in $\alpha$ is less significant.
$\Delta a_e $ has a mild anomaly at confidence level of $2.4~\sigma$, which might need a negative contribution from new physics. However, $\Delta \tilde{a}_e $ turns to be positive and the significance goes down to $1.6~\sigma$.
The measurement \cite{Morel:2020dww} placed $95 \%$ C.L. bounds to be $ \Delta \tilde{a}_{e} \in [-34 \times10^{-14}, 98\times 10^{-14}]$ and we will adopt this value in the later calculation.

\begin{figure}
	\centering	
	\includegraphics[width=0.4 \columnwidth]{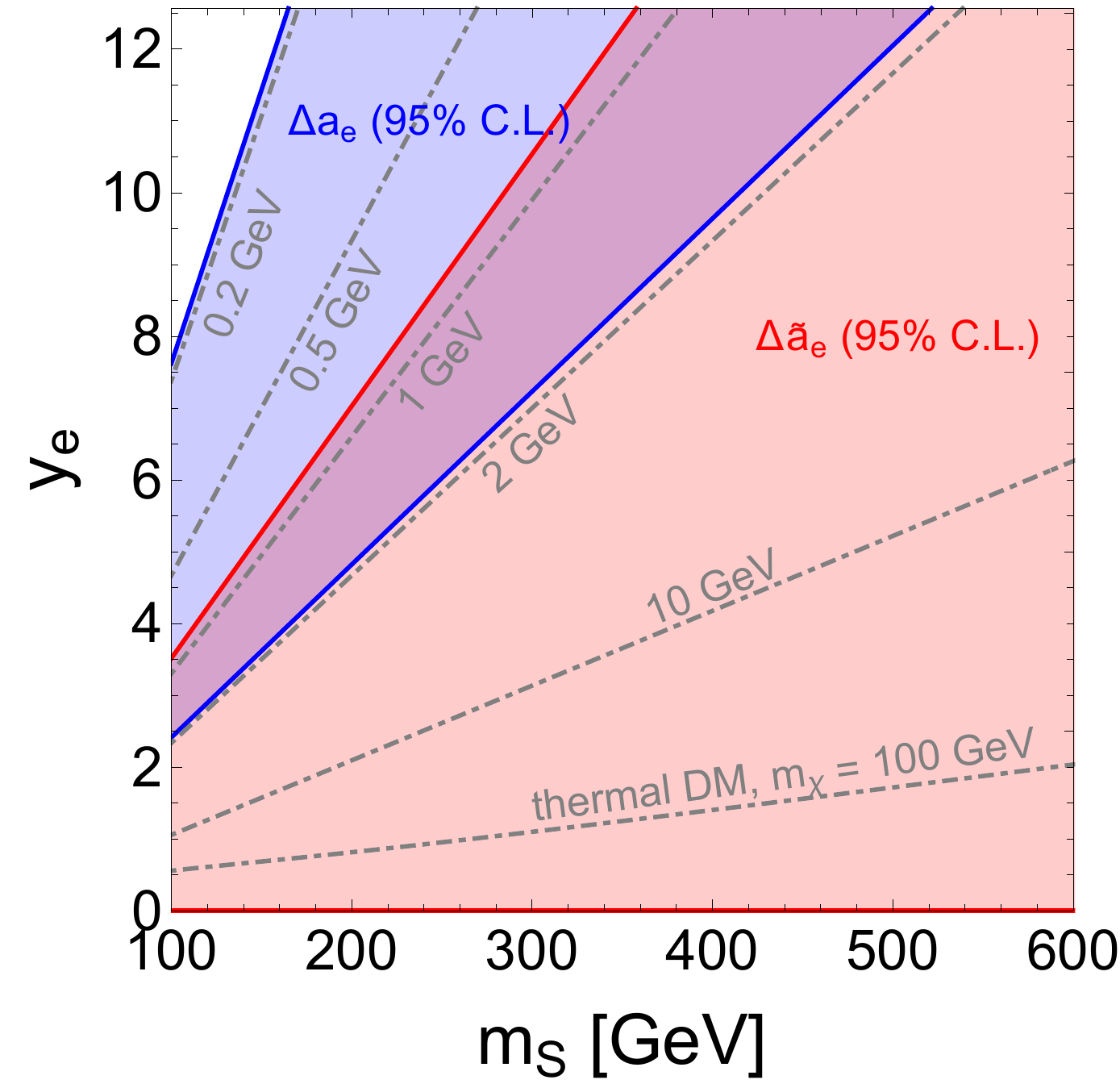}
	\caption{The interplay between $(g-2)_e$ and thermal DM requirements. In blue (red) shaded region, we show
		the range for $y_e$ which satisfies $\Delta a_e$ ($\Delta \tilde{a}_e$) at $95 \%$ C.L with $m_\chi \ll m_S$.
		The gray dotdashed lines shows the thermal $y_e^{\rm th}$ values with different $m_{\chi}$.
		For  both $\Delta a_e$ and $\Delta \tilde{a}_e$, $m_\chi$ between 1 GeV and 2 GeV can simultaneously explain $(g-2)_e$ and DM relic abundance.}
	\label{fig:muon-g-2}
\end{figure}

In our model, the contribution is always negative for the electron $g-2$. In Fig.~\ref{fig:muon-g-2}, We consider the $y_e$ range which
can fit to $\Delta a_e$ (blue shaded region) and $\Delta \tilde{a}_e$ (red shaded region) at $95 \%$ C.L., with a fixed $m_\chi $ 
which is much smaller than $m_S$. We fix $m_\chi = 1$ GeV for $(g-2)_e$ fit, which satisfies the condition $m_\chi \ll m_S$ easily.
For $g-2$ contribution in this limit, one can have the following expansion with small $x $ as
\begin{align}
	\Delta a_\ell \approx - \frac{y_\ell^2}{16 \pi^2} \frac{m_\ell^2}{m_S^2} \left(\frac{1}{6} - \frac{x}{3} +\mathcal{O}(x^2) \right).
\end{align}
Therefore, for $m_\chi$ considered in Fig.~\ref{fig:muon-g-2}, we see that $\Delta a_\ell$ is determined by $m_S$ and $y_\ell$ most of the time.
As a result, the blue and red shaded regions will not shift when changing to other $m_\chi$. 
On the other hand, we plot $y_e^{\rm th}$ with a given DM mass,
which can provide the correct DM relic abundance in Fig.~\ref{fig:muon-g-2}. They are plotted as dot-dashed gray lines with a
fixed $m_\chi$.  We can see that for $\Delta a_e$, DM mass between $0.2\sim2$ GeV is preferred and can satisfy DM relic abundance at the same time, while for $\Delta \tilde{a}_e$, DM mass larger than 1 GeV is required. The two measurements have a mutual region when DM mass is between 1 and 2 GeV.

The muon $g-2$ also has a long-standing discrepancy~\cite{Blum:2018mom, Bennett:2006fi}. Combination of the newest Fermilab and previous BNL measurements yields~\cite{Abi:2021gix}
\be
\Delta a_\mu = (2.51 \pm 0.59) \times 10^{-9},
\ee
corresponding to a significance of $4.2\sigma$, suggesting a positive contribution from the new physics.\footnote{There are debates on this ``excess''. A lattice group shows that there is no significant tension between the SM prediction and the recent FNAL experimental determination~\cite{Borsanyi:2020mff}.} Unfortunately, in this lepton portal model the contribution has a negative sign that it can not explain the anomaly. To incorporate the $\Delta a_\mu$ result, one has to add new ingredients beyond current model.

\section{Probing the model with gravitational waves}\label{sec:psgw}

In this section we investigate the possibility of probing the scalar sector via the GWs from a FOPT in the early universe.\footnote{The FOPT of such a singlet charged scalar was also studied in Ref.~\cite{Ahriche:2018rao}. Also see Refs.~\cite{Jiang:2015cwa,Chen:2019ebq,Cheng:2018ajh} for the FOPT triggered by a complex gauge singlet.} The first subsection is devoted to the discussion of FOPT, while the second subsection studies the GW detection limits.

\subsection{First-order phase transition}

The scalar potential in \Eq{eq:Lags} is 
\begin{equation*}
V(H, S)=\mu_H^2|H|^2+\mu_S ^2| S |^2+\lambda_H|H|^4+\lambda_S | S |^4+2\lambda_{H S }|H|^2| S |^2,
\end{equation*}
where 
\be
H=\frac{1}{\sqrt{2}}\begin{pmatrix}\sqrt{2}G^+\\ h+iG^0\end{pmatrix},\quad S=\frac{\phi+i\eta}{\sqrt{2}}.
\ee
In terms of the real components, we get
\be
V(h,\phi)=\frac{\mu_H^2}{2}h^2+\frac{\mu_S ^2}{2}\phi^2+\frac{\lambda_H}{4}h^4+\frac{\lambda_S }{4}\phi^4+\frac{\lambda_{H S }}{2}h^2\phi^2.
\ee
Here the quartic coefficients should satisfy
\be\label{bounded_below}
\lambda_H>0,\quad \lambda_S >0,\quad \sqrt{\lambda_H\lambda_S }+\lambda_{H S }>0,
\ee
to ensure the potential is bounded below.

Since $ S $ has electric charge number $-1$, it cannot develop a vacuum expectation value (VEV) at zero temperature. Hence the vacuum configuration at zero temperature is along the Higgs direction, i.e. $(\ave{h},\ave{\phi})=(v,0)$. This means that the Higgs-relevant coefficients have been fixed by the collider measurements, 
\be
\mu_H^2=-\frac{m_h^2}{2},\quad \lambda_H=\frac{m_h^2}{2v^2},
\ee
where $m_h=125$ GeV and $v=246$ GeV, leaving only three free parameters in the scalar potential. If $\mu_S ^2$ is positive, then the vacuum configuration $(v,0)$ is trivially achieved, because along the $\phi$ direction there is no local minimum. However, if $\mu_S^2<0$, then the $\phi$ direction also has a local minimum $w=\sqrt{-\mu_S^2/\lambda_S }$. In this case, to make sure $(v,0)$ is a minimum but not a saddle point, the coefficients must satisfy
\be
\lambda_H\mu_S ^2>\lambda_{H S }\mu_H^2,
\ee
according to the Hessian matrix~\cite{Bian:2019kmg}. In case of $\lambda_S \mu_H^2>\lambda_{H S }\mu_S ^2$, $(0,w)$ is also a local minimum, and a further condition
\be
V(v,0)=-\frac{\mu_H^4}{4\lambda_H}<V(0,w)=-\frac{\mu_S ^4}{4\lambda_S },
\ee
is required to ensure $(v,0)$ is the global minimum. The mass of $S$ is given by $m_S ^2=\mu_S ^2+\lambda_{HS}v^2$. Although the collider experiments have set a bound for $m_S$ (which also depends on $m_\chi$), they cannot probe $\mu_S ^2$ directly and neither the sign of $\mu_S^2$. If $\mu_S^2<0$, the potential might trigger a FOPT in the early universe, and this opens a new detection scenario for this kind of DM models: the FOPT GWs.

At finite temperature, the scalar potential is modified by the thermal correction. Taking the leading gauge invariant $T^2$ terms~\cite{Dolan:1973qd,Braaten:1989kk}, the thermal potential is
\be\label{VT_sim}
V_T(h,\phi,T)\approx\frac{\mu_H^2+c_hT^2}{2}h^2+\frac{\mu_S ^2+c_\phi T^2}{2}\phi^2+\frac{\lambda_H}{4}h^4+\frac{\lambda_S }{4}\phi^4+\frac{\lambda_{H S }}{2}h^2\phi^2,
\ee
where
\be\label{ch_cphi}
c_h=\frac{3g^2+g'^2}{16}+\frac{y_t^2}{4}+\frac{\lambda_H}{2}+\frac{\lambda_{HS}}{6},\quad c_\phi=\frac{g'^2}{4}+\frac{\lambda_S}{3}+\frac{\lambda_{H S }}{3}.
\ee
The necessary condition for a FOPT is the existence of a critical temperature $T_c$ at which the system has two energetically degenerate vacua $(v_c,0)$ and $(0,w_c)$. For \Eq{VT_sim}, this requires~\cite{Bian:2019kmg}
\be\label{analytical}
\frac{c_\phi}{c_h}<\frac{\mu_S^2}{\mu_H^2}<\frac{\sqrt{\lambda_S}}{\sqrt{\lambda_H}}<\frac{\lambda_{HS}}{\lambda_H},
\ee
and the critical temperature and VEVs are respectively
\be
T_c=\sqrt{\frac{\mu_H^2\sqrt{\lambda_S}-\mu_S^2\sqrt{\lambda_H}}{c_\phi\sqrt{\lambda_H}-c_h\sqrt{\lambda_S}}},\quad
v_c=\sqrt{\frac{c_h\mu_S ^2-c_\phi\mu_H^2}{c_\phi\lambda_H-c_h\sqrt{\lambda_H\lambda_S}}}.
\ee
Below $T_c$, the Higgs-direction minimum becomes the lower one and the universe starts to decay to it from the $\phi$-direction minimum. The decay rate per unit volume reads $\Gamma(T)\sim T^4e^{-S_3/T}$, where $S_3$ is the classical action for the $O(3)$ symmetric bounce solution~\cite{Linde:1981zj}. The nucleation temperature $T_n$ is defined by the equality of the nucleation rate per Hubble volume and the universe expansion rate, i.e. $\Gamma(T_n)=H^4(T_n)$. For a radiation-dominated universe and a FOPT happening around the EW scale, $T_n$ can be solved by~\cite{Quiros:1999jp}
\be\label{FOPT}
S_3/T_n\sim140.
\ee
\Eq{FOPT} is treated as the sufficient condition for a FOPT in this article.

\begin{figure}
\centering
\includegraphics[width=0.4\columnwidth]{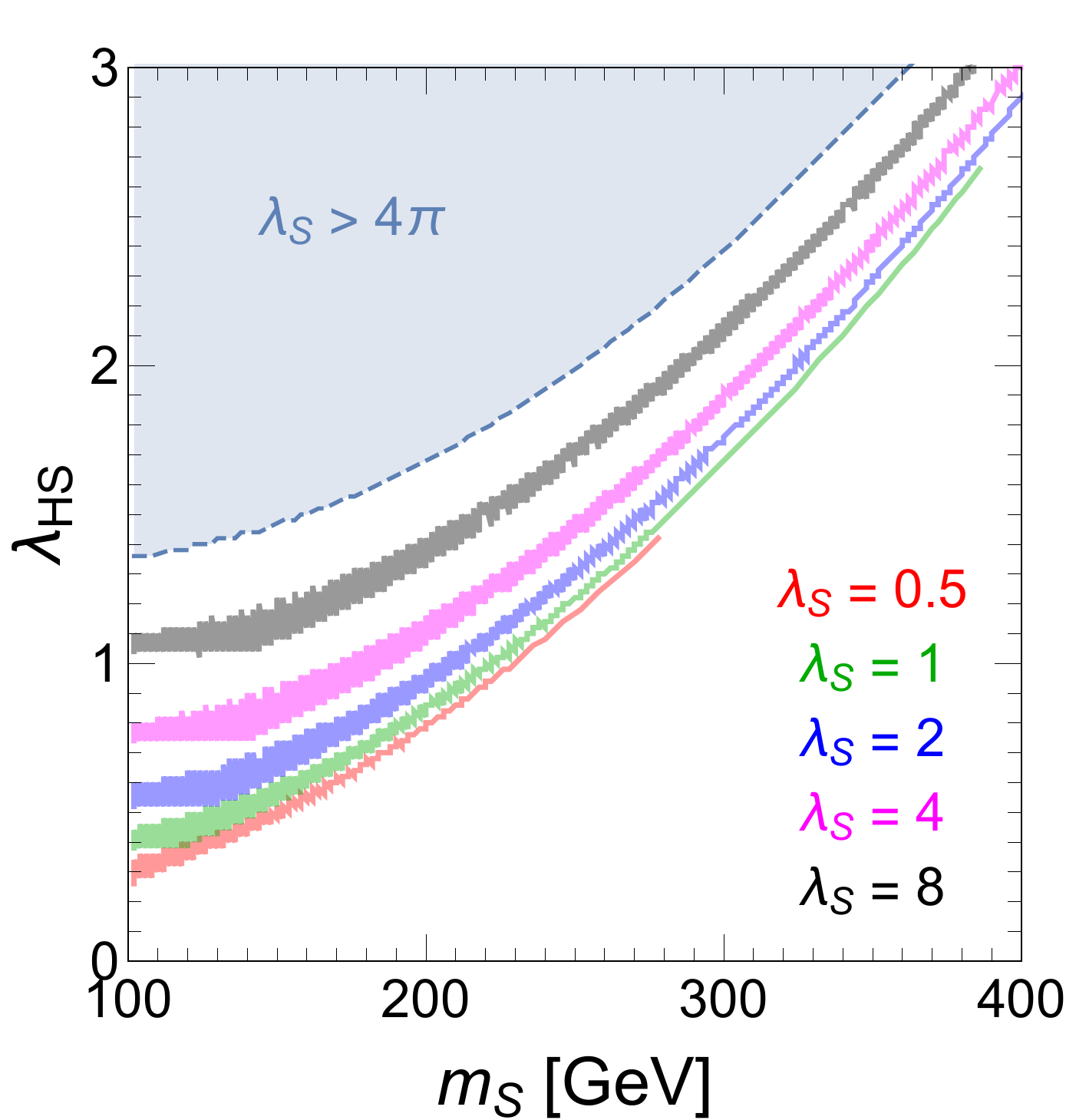}
\caption{The parameter space allowed by a FOPT. Different colors correspond to different $\lambda_S$ values.}
\label{fig:EWPT}
\end{figure}

We use the Python package {\tt cosmoTransition}~\cite{Wainwright:2011kj} to calculate the bounce solution and hence $T_n$ for the potential in \Eq{VT_sim}. As discussed before, after taking into account the Higgs mass and VEV, in the scalar potential there are only three free parameters, which we choose to be $\lambda_S$, $m_S$ and $\lambda_{HS}$. For a fixed $\lambda_S$, the parameter space allowed by the FOPT can be projected to the $m_S$-$\lambda_{HS}$ plane. We plot the parameter space for different $\lambda_S$ as shaded areas in Fig.~\ref{fig:EWPT}. The shapes of those areas can be well understood by the analytical relation \Eq{analytical}. A sizable $\lambda_{HS}$ is preferred by FOPT, because the phase transition is triggered by the potential barrier induced by the $\lambda_{HS}h^2S^2/2$ term.

\subsection{Gravitational waves}

Stochastic GWs are produced during a FOPT via bubble collision~\cite{Huber:2008hg,Di:2020nny}, sound waves in the plasma~\cite{Hindmarsh:2015qta} and the magneto-hydrodynamics turbulence~\cite{Binetruy:2012ze,Caprini:2009yp}. The expanding bubble wall only accelerates for a short time before it reaches its final velocity $v_b<1$. Therefore, only a tiny fraction of FOPT energy deposits in the bubble shell, and the bubble collision contribution to the GWs is negligible. Instead, most of the phase transition energy is pumped into the surrounding fluid shells, making sound waves the dominant contribution to FOPT GWs~\cite{Ellis:2018mja}. However, the sound wave period only lasts for a finite time, after which the energy in the bulk fluid will cause the turbulence, which is another source for GWs~\cite{Ellis:2020awk}. Consequently, the GW spectrum today can be expressed as
\be
\Omega_{\rm GW}(f)=\Omega_{\rm sw}(f)+\Omega_{\rm turb}(f),
\ee
where $f$ is the frequency, the subscripts ``sw'' and ``turb'' denote sound waves and turbulence respectively, and $\Omega_{\rm GW}$ is the ratio of GW energy density to the critical energy of the current universe, i.e.
\be\label{GW}
\Omega_{\rm GW}(f)=\frac{1}{\rho_c}\frac{\rho_{\rm GW}}{d\ln f},
\ee
where $\rho_c=3H_0^2/(8\pi G)$, with $H_0$ being the Hubble constant today.

The sound wave and turbulence spectra can be expressed as functions of two FOPT parameters~\cite{Grojean:2006bp,Caprini:2015zlo,Caprini:2019egz},
\be
\alpha=\frac{1}{g_*\pi^2T_n^4/30}\left(T\frac{\partial\Delta V_T}{\partial T}-\Delta V_T\right)\Big|_{T_n};\quad \beta/H=T_n\frac{d(S_{3}/T)}{dT}\Big|_{T_n},
\ee
where $\Delta V_T$ denotes the (negative) effective potential difference between the true and false vacua, and $g_*\sim100$ is the number of relativistic degrees of freedom during FOPT.\footnote{It is suggested that the $\alpha$ and $\beta/H$ parameters should be calculated at the percolation temperature $T_p$~\cite{Megevand:2016lpr,Kobakhidze:2017mru,Ellis:2018mja,Ellis:2020awk,Wang:2020jrd}. However, we have checked that for our FOPT scenario $\alpha\lesssim1$, therefore the supercooling effect is not prominent and $T_n\approx T_p$ is a good approximation.} Namely, $\alpha$ is the ratio of FOPT latent heat to the radiation energy, while $\beta/H$ is the inverse ratio of FOPT duration to the universe expansion time scale. Numerically, the sound waves spectrum is~\cite{Hindmarsh:2015qta}
\be\label{sw}
\Omega_{\rm sw}(f)h^2=2.65\times10^{-6}\frac{1}{\beta/H}\left(\frac{\kappa_v\alpha}{1+\alpha}\right)^2\left(\frac{g_*}{100}\right)^{-1/3}v_b\left(\frac{f}{f_{\rm sw}}\right)^3\left(\frac{7}{4+3(f/f_{\rm sw})^2}\right)^{7/2},
\ee
where
\be\label{sw_f}
f_{\rm sw}=1.9\times10^{-2}~{\rm mHz}\times\frac{\beta/H}{v_b}\left(\frac{T_n}{100~{\rm GeV}}\right)\left(\frac{g_*}{100}\right)^{1/6}.
\ee
To take into account the finite duration of the sound wave period, we make the replacement $\Omega_{\rm sw}(f)\to \Omega_{\rm sw}(f)H(T_n)\tau_{\rm sw}$ in \Eq{sw}~\cite{Ellis:2020awk,Wang:2020jrd}, where
\be
\tau_{\rm sw}={\rm min}\left\{\frac{1}{H(T_n)},\frac{v_b(8\pi)^{1/3}}{\beta\bar U_f}\right\},\quad \bar U_f=\sqrt{\frac{3}{4}\frac{\kappa_v\alpha}{1+\alpha}}.
\ee
A more accurate treatment for the sound wave cutoff factor can be found in Ref.~\cite{Guo:2020grp}.

The turbulence spectrum is~\cite{Binetruy:2012ze,Caprini:2009yp}
\be\label{turb}
\Omega_{\rm turb}(f)h^2=3.35\times10^{-4}\frac{v_b}{\beta/H}\left(\frac{\kappa_{\rm turb}\alpha}{1+\alpha}\right)^{3/2}\left(\frac{g_*}{100}\right)^{-1/3}S_{\rm turb}(f),
\ee
where
\be
S_{\rm turb}(f)=\frac{(f/f_{\rm turb})^3}{\left[1+(f/f_{\rm turb})\right]^{11/3}(1+8\pi f/h_*)},\quad h_*=16.5\times10^{-3}~{\rm mHz}\left(\frac{T_n}{100~{\rm GeV}}\right)\left(\frac{g_*}{100}\right)^{1/6},
\ee
and
\be
f_{\rm turb}=2.7\times10^{-2}~{\rm mHz}\times\frac{\beta/H}{v_b}\left(\frac{T_n}{100~{\rm GeV}}\right)\left(\frac{g_*}{100}\right)^{1/6}.
\ee
The factor $\kappa_{v}$ and $\kappa_{\rm turb}$ in \Eq{sw} and \Eq{turb} are the fraction of FOPT energy that is transformed to bulk motion/turbulence, respectively. We have adopted $\kappa_{\rm turb}=0.05\kappa_v$, and $\kappa_v$ is extracted from the numerical function of Ref.~\cite{Espinosa:2010hh}. $v_b=0.6$ is used as a benchmark in our study, although it might be calculated using the hydrodynamics.\footnote{See Refs.~\cite{Konstandin:2014zta,Cline:2021iff} for the $v_b$ determination in the real singlet extended SM.}

\begin{figure}
\centering
\includegraphics[width=0.4\columnwidth]{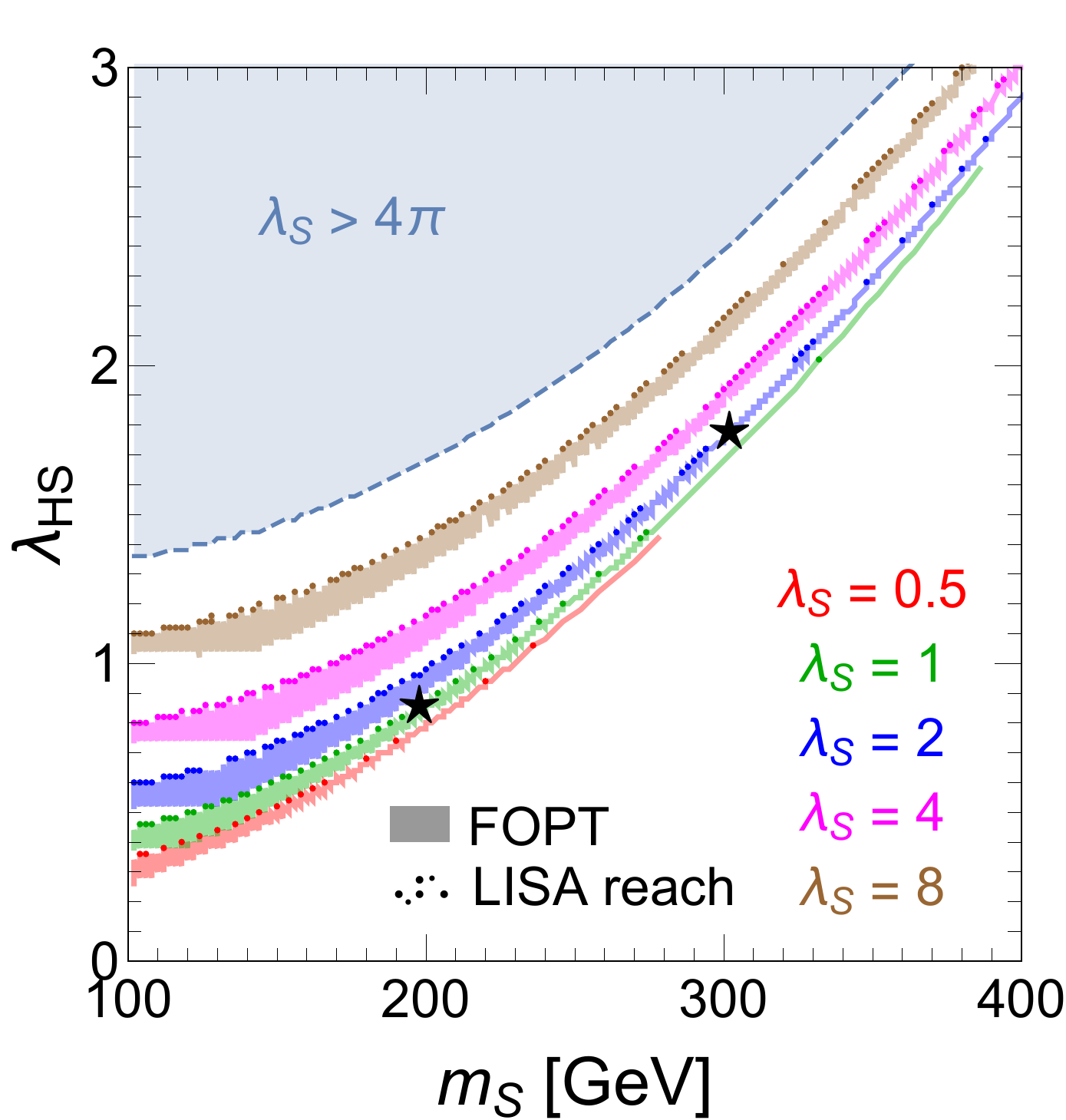}\qquad\qquad
\includegraphics[width=0.4\columnwidth]{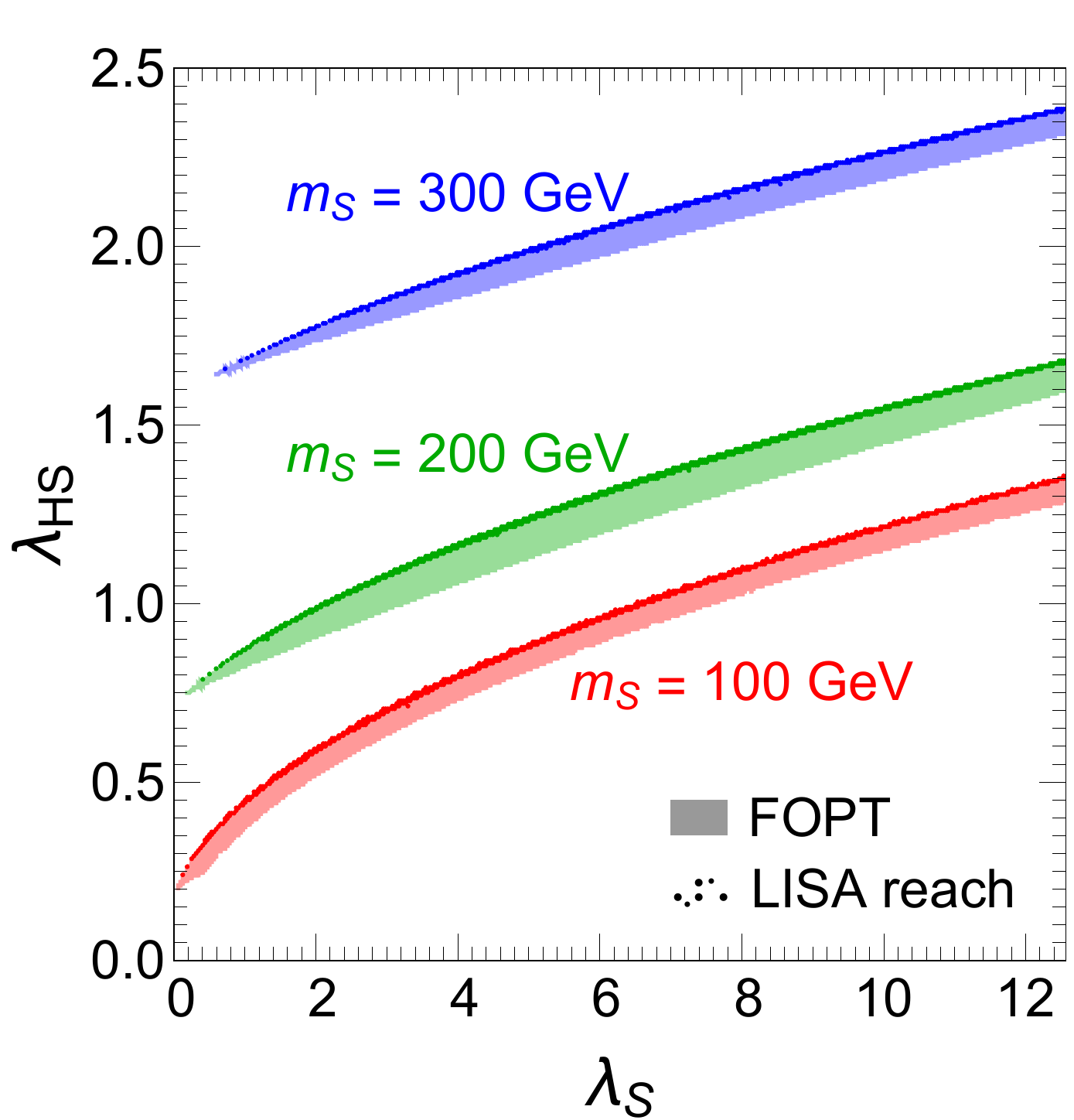}
\includegraphics[width=0.435\columnwidth]{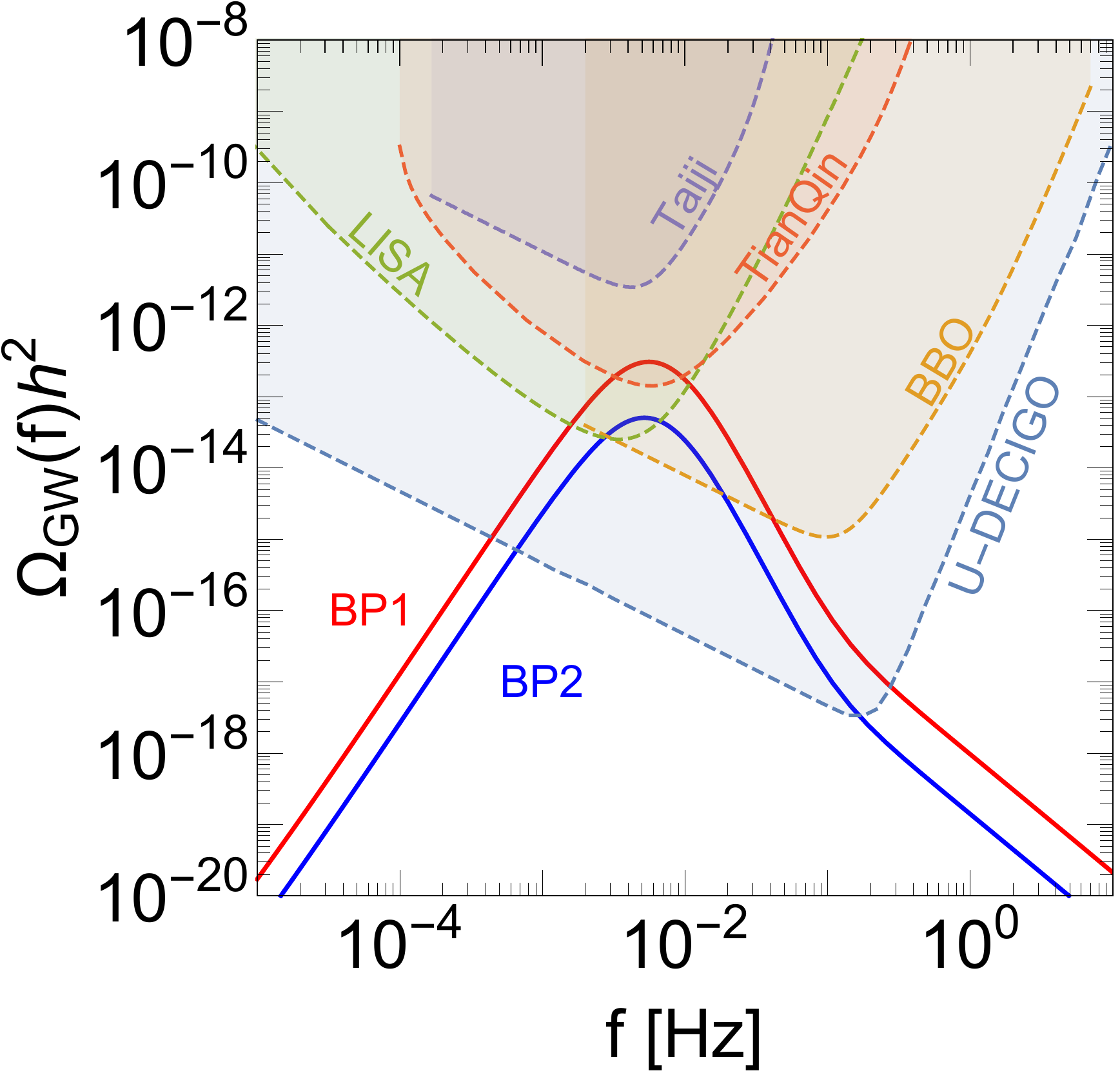}
\caption{Top: the parameter space that triggers FOPT and can be probed by LISA, projected to different 2-dimension planes. Bottom: the GW spectrum from two benchmark points which are labeled as stars in the top left panel.}
\label{fig:EWPT_GWs}
\end{figure}

The GWs from a FOPT around EW scale might be probed by the next generation space-based laser interferometers such as LISA~\cite{Audley:2017drz}, BBO~\cite{Crowder:2005nr}, TianQin~\cite{Luo:2015ght,Hu:2017yoc}, Taiji~\cite{Hu:2017mde,Guo:2018npi} and DECIGO~\cite{Kawamura:2011zz,Kawamura:2006up}. The detectability of GW signals can be characterized by the signal-to-noise ratio (SNR). Taking the LISA detector as an example, the SNR is defined as~\cite{Caprini:2015zlo}
\be
{\rm SNR}=\sqrt{\mT\int_{f_{\rm min}}^{f_{\rm max}} df\left(\frac{\Omega_{\rm GW}(f)}{\Omega_{\rm LISA}(f)}\right)^2},
\ee
where $\Omega_{\rm LISA}$ is the sensitive curve of the LISA detector~\cite{Caprini:2015zlo} (in which we take the C1 configuration as a benchmark), and $\mT=9.46 \times 10^7$~s is the data-taking duration~\cite{Caprini:2019egz}. We adopt ${\rm SNR}=10$ as the detectable threshold, and found that for a fixed $\lambda_S$, only a narrow band of parameter space in Fig.~\ref{fig:EWPT} can be probed. The parameter space that can be probed by LISA is plotted in the top panel of Fig.~\ref{fig:EWPT_GWs}, where we show the results both in the $m_S$-$\lambda_{HS}$ plane (for fixed $\lambda_S$) and the $\lambda_S$-$\lambda_{HS}$ plane (for fixed $m_S$). In the bottom panel of the same figure we plot the GW spectrum for two benchmark points:
\be\label{BPs_1}\begin{split}
{\rm BP1}:\quad&m_S=198~{\rm GeV},~ \lambda_{HS}=0.86,~ \lambda_S=1,\\
&T_n=76.6~{\rm GeV},~ v_n=189~{\rm GeV},~ w_n=94.9~{\rm GeV},~ \alpha=0.0834,~\beta/H=228;\\ 
{\rm BP2}:\quad&m_S=302~{\rm GeV},~ \lambda_{HS}=1.78,~ \lambda_S=2,\\
&T_n=91.8~{\rm GeV},~ v_n=123~{\rm GeV},~ w_n=53.1~{\rm GeV},~ \alpha=0.0248,~\beta/H=177;\\ 
\end{split}\ee
where $(0,w_n)$ and $(v_n,0)$ are respectively the old and new vacua at the nucleation temperature $T_n$.
 
\section{The interplay between phase transition and particle searches}\label{sec:interplay}
 
In this section, we revisit the parameter space for $\lambda_{HS}$ from GW study and the corresponding constraints from the collider studies. We focus on how the two different types of experiments can be complimentary with each other.
For GW experiment, we focus on the LISA detectable parameter space in the left panel of Fig.~\ref{fig:EWPT_GWs}. Clearly, the scalar $S$ self-interaction coupling $\lambda_S$ can affect the LISA detectable parameter space. Therefore, we vary $\lambda_S$ between $0$ and $4\pi$ to obtain the entire GW detectable region for a FOPT.
 
\subsection{Interplay with $pp\to S^+S^-$ at the LHC}

\begin{figure}
\centering	
\includegraphics[width=0.4 \columnwidth]{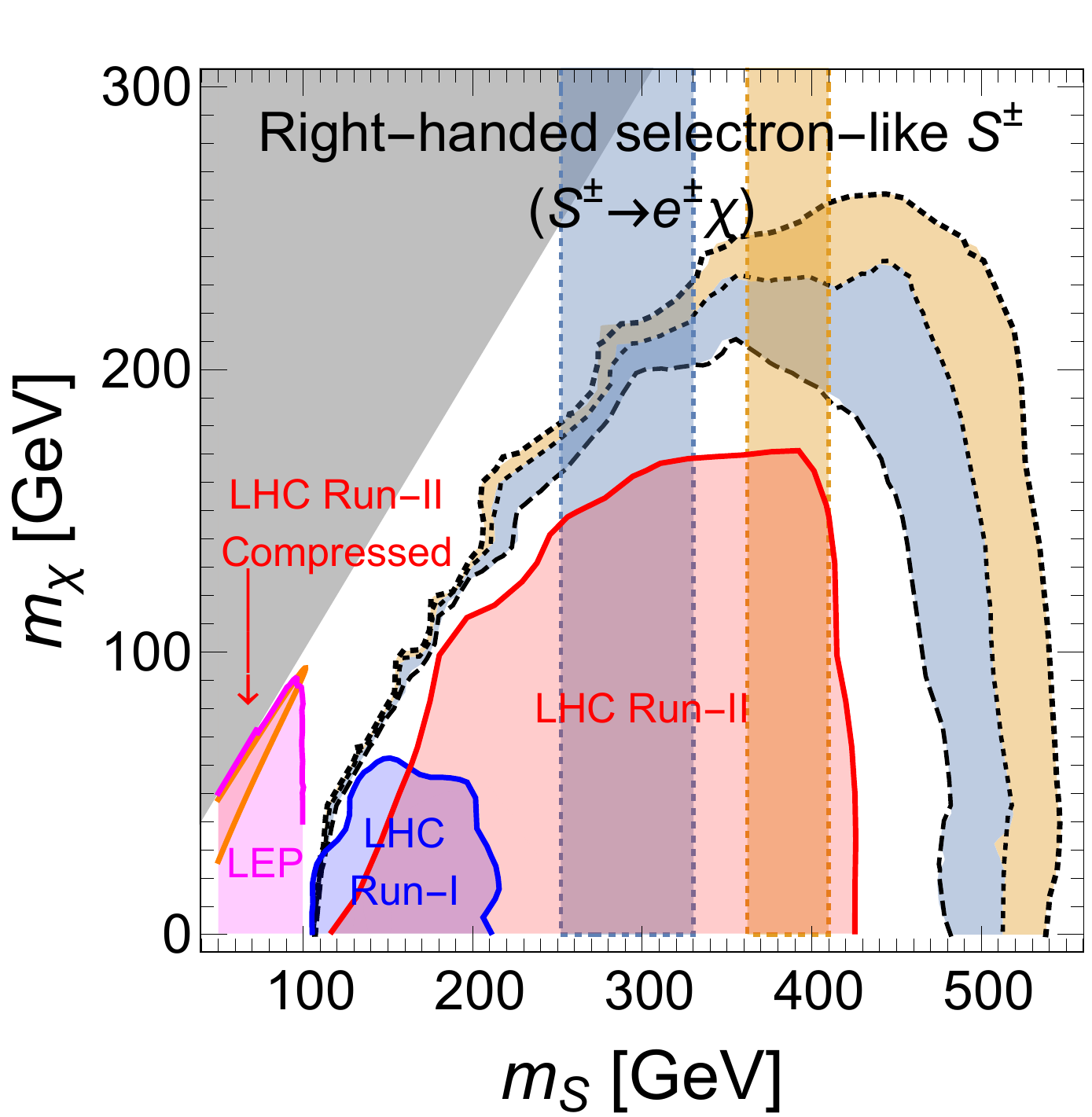} \qquad
\includegraphics[width=0.4 \columnwidth]{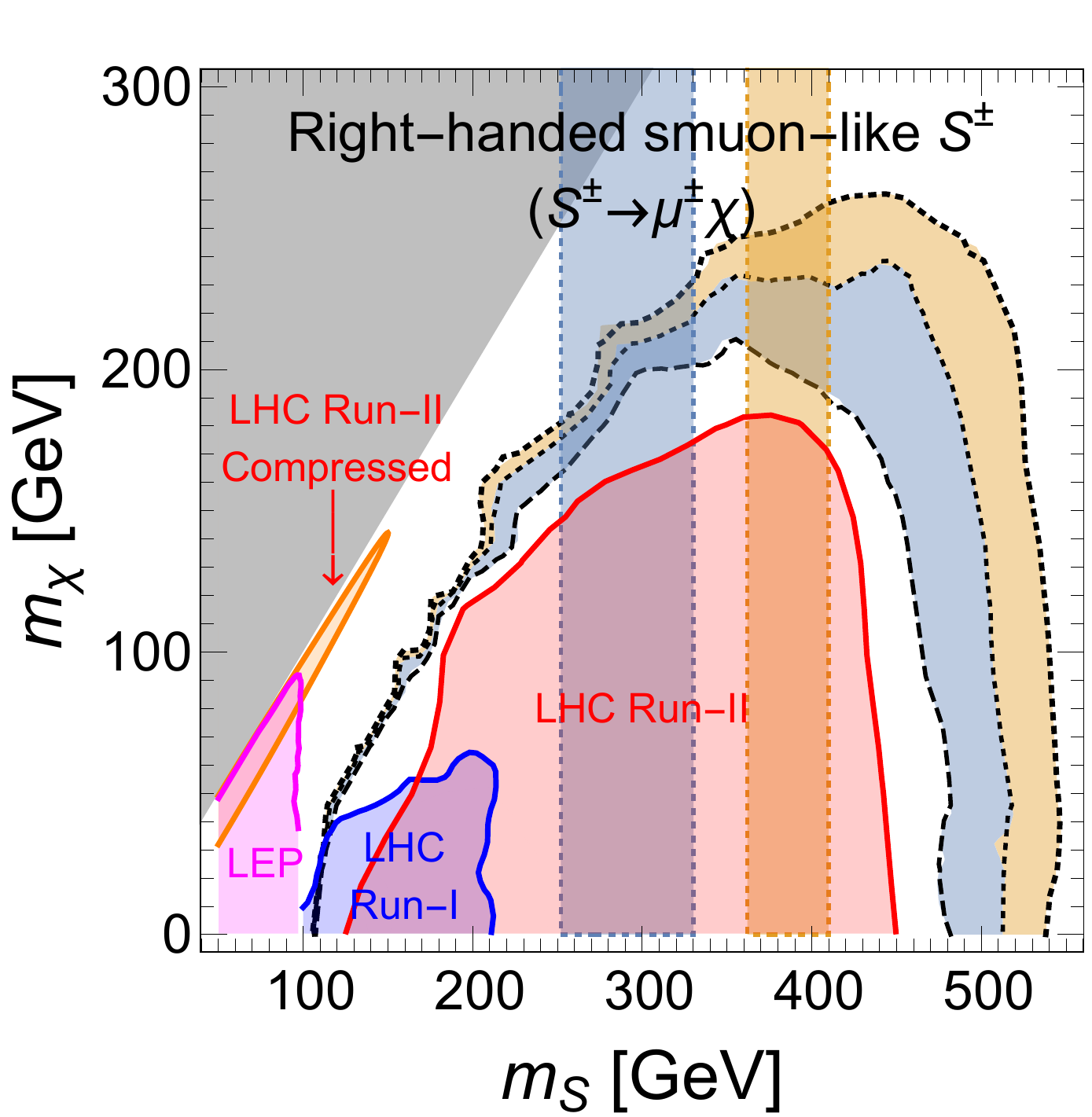}
\caption{The interplay between gravitational wave detection and LHC searches. The collider constrains and projections are the same as those in Fig.~\ref{fig:slepton-constraints}. The light blue (orange) shaded region corresponds to $\lambda_{HS}=2$ (3), with the vertical boxed boundary regions being the LISA-detectable parameter space, while the irregular boundary regions being enhanced part of the LHC projections when including the $gg\to h^*\to S^+S^-$ contribution.}
\label{fig:ms-mchi-GW}
\end{figure}

The $S^\pm$ particles can be pair produced at the LHC via Drell-Yan and gluon-gluon fusion processes, and the current constrains and future projections at the 14 TeV LHC has been studied in Section~\ref{sec:pp_SS}. As shown in Fig.~\ref{fig:slepton-constraints}, a non-zero $\lambda_{HS}$ can visibly enhance the probe limit in parameter space due to the $gg\to h^*\to S^+S^-$ contribution to the signal events. On the other hand, a sizable $\lambda_{HS}$ may trigger a FOPT and hence give detectable GW signals as well.

The interplay between the LHC and the future LISA experiments is plotted in Fig.~\ref{fig:ms-mchi-GW}. The light blue (orange) shaded region corresponds to $\lambda_{HS}=2$ (3), with the vertical boxed boundary regions being the LISA-detectable parameter space, while the irregular boundary regions being the enhanced LHC projections when including the $gg\to h^*\to S^+S^-$ contribution. This is because the GW signals is independent of the DM mass $m_\chi$. For a given $\lambda_{HS}$, there is a set of upper and lower bounds for $m_S$ in the LISA-detectable region. The enhanced parts of the LHC probe region due to $\lambda_{HS}$ are also shown in the figure. We see that the LHC and GW experiments mainly serve as complementary approaches to probe the DM parameter space; while they also have some intersections, which can be used to identify the origin of the excess (if found).

\subsection{Higgs precision measurement at the future $e^+e^-$ colliders}

\begin{figure}
	\centering	
	\includegraphics[width=0.32 \columnwidth]{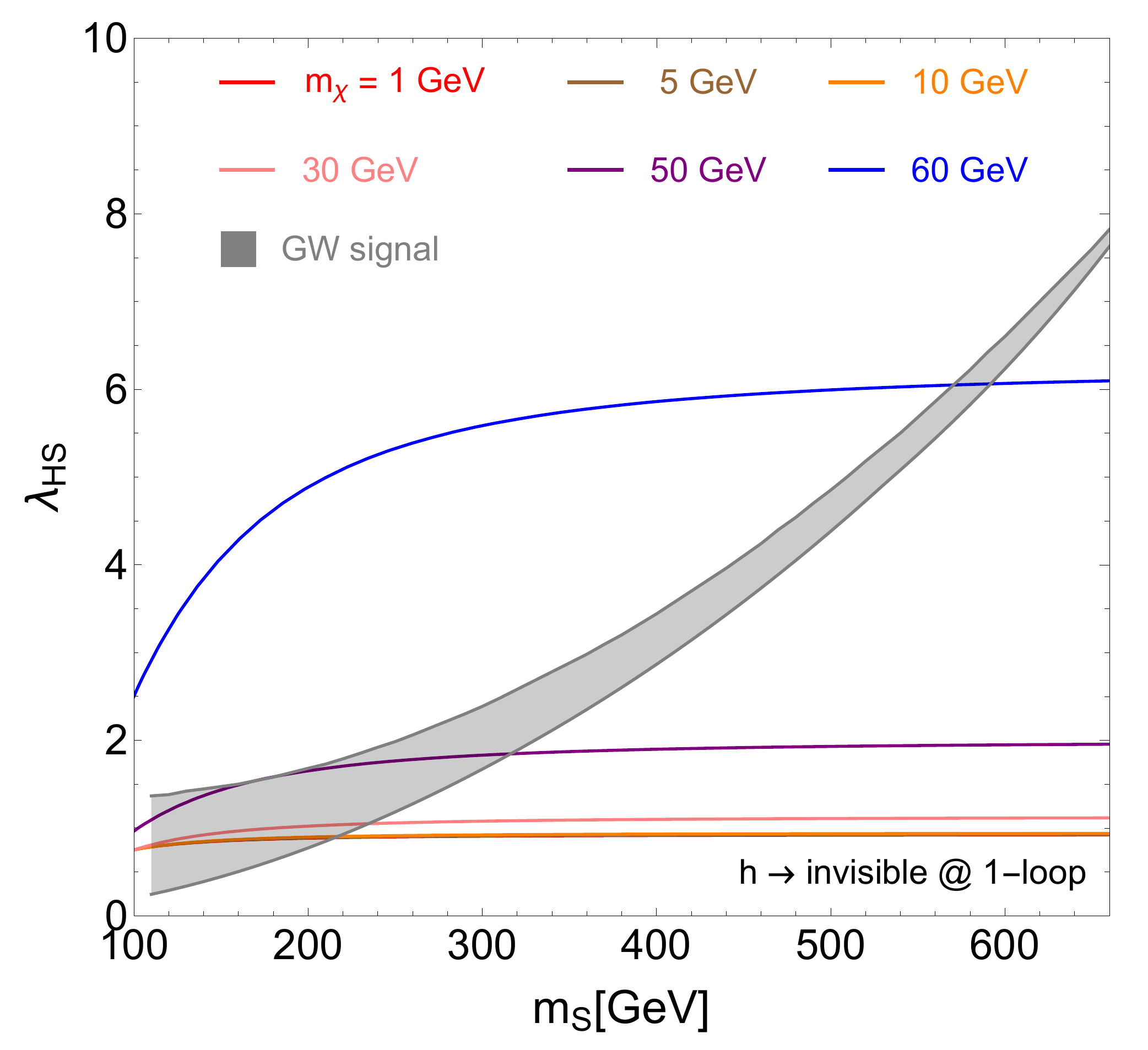}
	\includegraphics[width=0.32 \columnwidth]{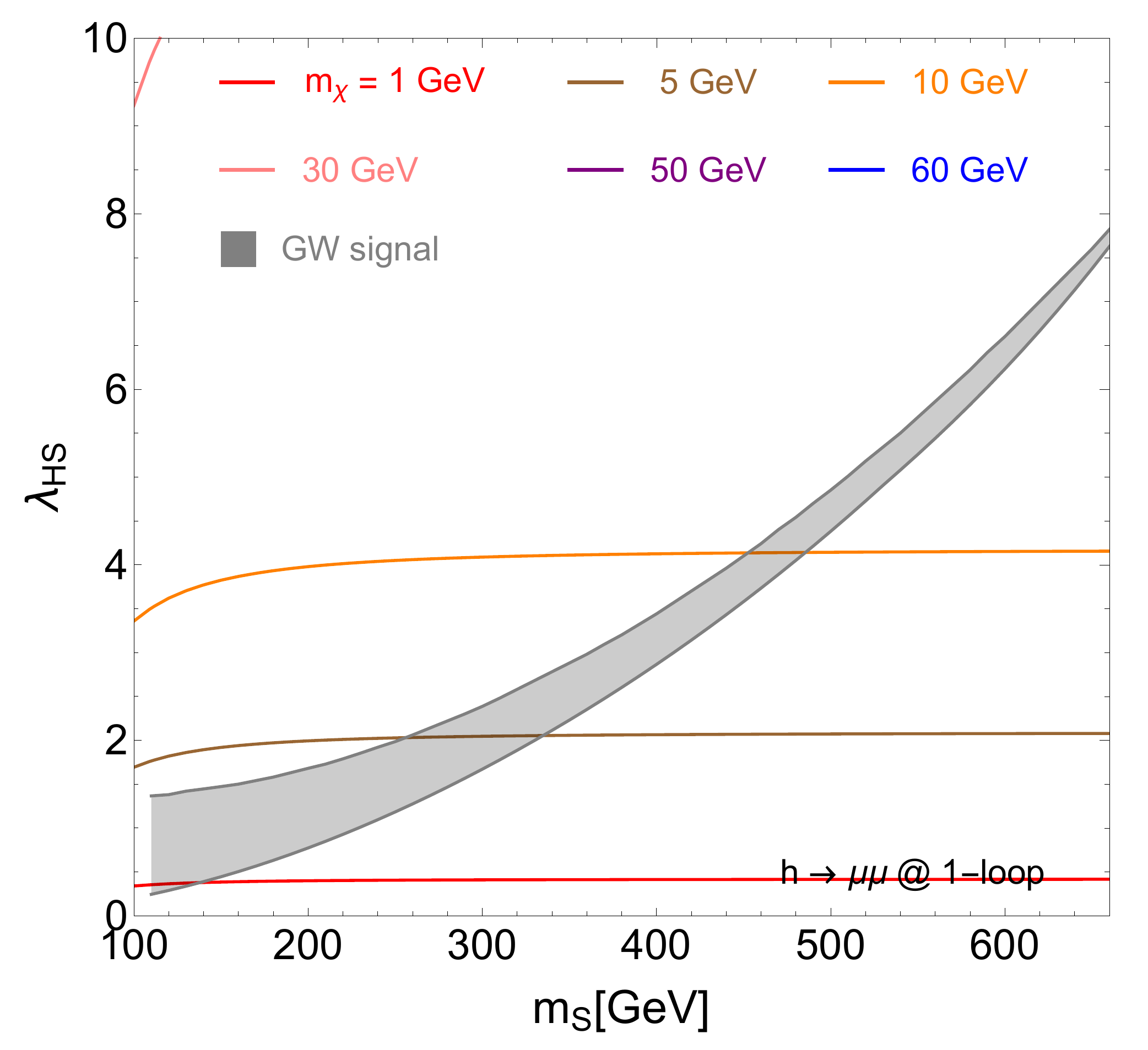}
	\includegraphics[width=0.32 \columnwidth]{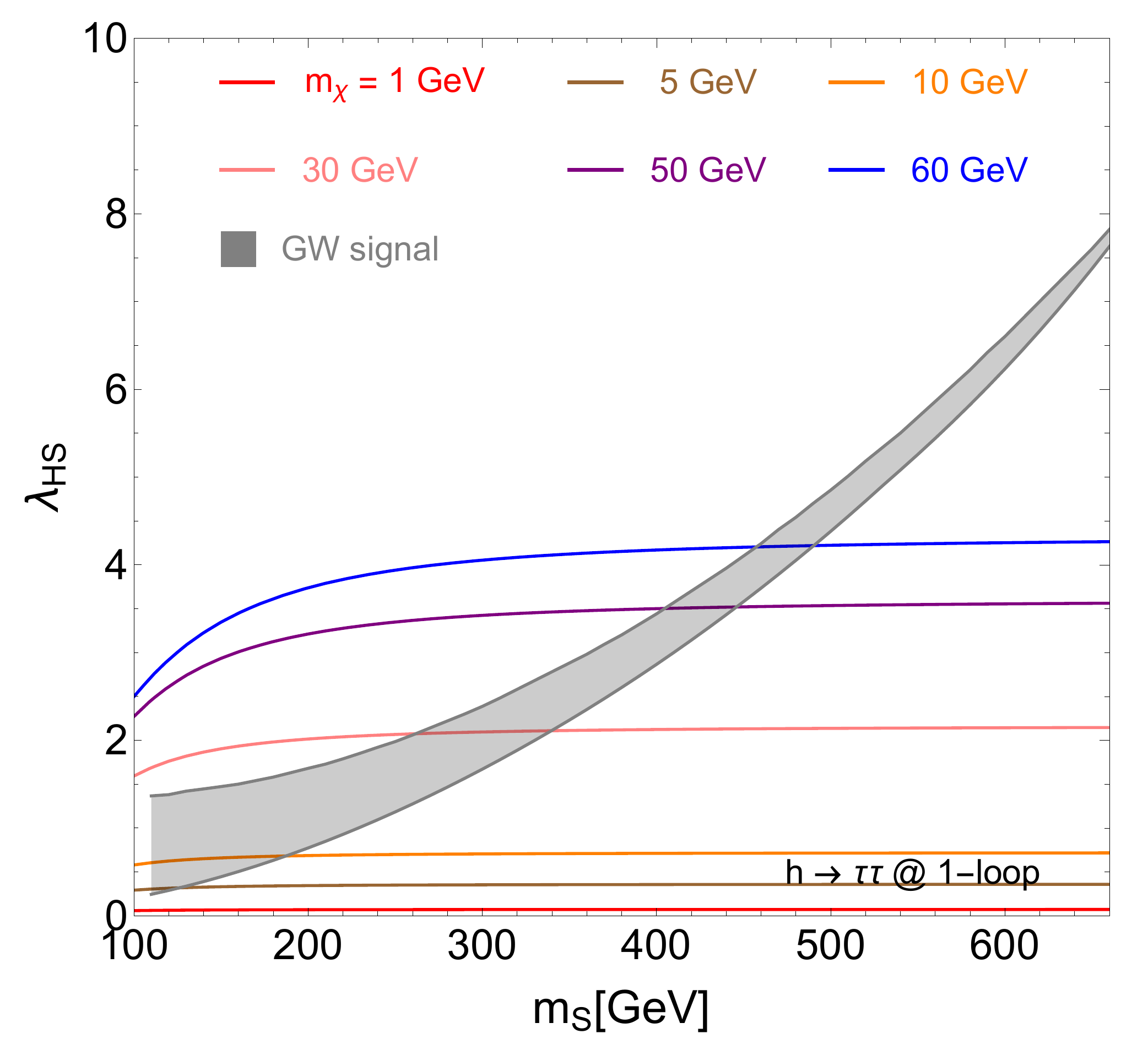}
	\caption{The interplay between gravitational wave detection and future $e^+e^-$ collider searches. 
		The gray shaded region is the LISA detectable parameter space, varying $\lambda_S$ from 0 to $4 \pi$.
		From left to right, we show the sensitivities for $\lambda_{HS}$ from future CEPC precision measurements,
		based on invisible Higgs decay branching ratio $\Br(h\to{\rm inv}) = 0.3\%$, Higgs leptonic coupling precision reaches
		$\delta \kappa_\mu < 8.7\%$ and $\delta \kappa_\tau <1.5\%$. }
	\label{fig:h-1loop-GW}
\end{figure}

The collider constraints on $\lambda_{HS}$ have to be related with SM Higgs. The constraint from exotic Higgs decay
is less sensitive compared to the Higgs 1-loop coupling as shown in the previous section.
The 1-loop induced Higgs couplings include the coupling to $\chi\chi  $ and $\ell^+\ell^-$.
The former can be revealed by the Higgs invisible decay branching ratio, for example we consider the future
sensitivity from CEPC $\Br(h\to{\rm inv}) = 0.3\%$.
The latter can be revealed by the Higgs precision measurements at CEPC with relative precision of couplings 
$\delta \kappa_\mu < 8.7\%$ and $\delta \kappa_\tau <1.5\%$. 
In Fig.~\ref{fig:h-1loop-GW}, we take the DM mass $m_\chi = 1, ~ 5, ~ 10, ~ 30 , ~ 50 , ~60$ GeV respectively to show its effect on the sensitivities for $\lambda_{HS}$. For a fixed DM mass, the corresponding colored line shows the maximum allowed $\lambda_{HS}$ from the future $e^+ e^-$ collider searches. In general, the exclusion power is better for light DM mass $m_\chi$.

In the left panel of Fig.~\ref{fig:h-1loop-GW}, one can see that for $m_\chi < 40 $ GeV the constraints on $\lambda_{HS}$ are quite similar. The reason is that the 1-loop induced coupling is proportional to 
$y^2_\ell \lambda_{HS} m_\chi / m_S^2$ for large $m_S$. At the same time, the annihilation cross section
is proportional to $y^4_\ell m_\chi^2 / m_S^4$ which requires this combination to be a constant to satisfy
the relic abundance. Therefore, the limits on $\lambda_{HS}$ from Higgs invisible decay is a constant. 
The colored lines in the left panel do show this feature, except when $m_S$ is too close to the Higgs mass and
the expansion on large $m_S$ is not valid anymore, the sensitivity on $\lambda_{HS}$ changes slightly. 
For larger $m_\chi$, the sensitivity of $\lambda_{HS}$
is downgraded because the phase space suppression in the  $h \to \chi \chi$ decay.
It is worth to mention that the sensitivity from Higgs invisible decay works equally good for all three flavors 
of lepton portals. This search can test most of the LISA detectable parameter regions for $m_\chi < m_h/2$.

In the right panel of Fig.~\ref{fig:h-1loop-GW}, the limits from $\delta \kappa_\tau <1.5\%$ are plotted for
different DM mass. For large $m_S$, we can see that the constraints on $\lambda_{HS}$ are proportional to 
$m_\chi$. The reason is that the one-loop induced Higgs coupling is roughly proportional to 
$y^2_\ell\lambda_{HS} m_\ell/m_S^2$ for large $m_S$ expansion. Since the relic abundance fix the 
combination $y^2_\ell m_\chi / m_S^2$ to be constant, the sensitivity for $\lambda_{HS}$ from Higgs 
precision measurement is proportional to $m_\chi$. Different from Higgs invisible branching ratio, there is no phase
space suppression for $m_\chi \sim m_h/2$. One can see that for $m_\chi \lesssim 20$ GeV, the Higgs-tau coupling precision measurement is the most sensitive among the three panels in the figure, while for the intermediate mass $m_\chi $ between $ 20$ to $\sim 50 $ GeV the Higgs invisible branching ratio measurement is better. For $m_\chi $ close to $m_h/2$, the Higgs-tau coupling measurement becomes better again due to the phase space suppression in the Higgs invisible decay.

In the middle panel of Fig.~\ref{fig:h-1loop-GW}, we show the limits from $\delta \kappa_\mu < 8.7\%$.
The results from muon coupling measurements are fully analogous to tau coupling. The sensitivity is worse
by a constant factor from $\delta \kappa_\tau /\delta \kappa_\mu$, reflecting the fact that more taus are produced
due to larger Higgs-tau coupling.

Finally, it should be mentioned that the limits on $\lambda_{HS}$ from $\Br(h \to \gamma \gamma)$ and $\sigma(Zh)$ 
are also very powerful as shown in section \ref{sec:1loop} and are able to exclude most of the parameter space for GW
detection \cite{Huang:2016cjm}. Such constraints are independent of the DM Yukawa coupling
$y_\ell$ and therefore, are complementary with the limits from $h$ invisible and leptonic decays.
 
\section{Conclusions}\label{sec:conclusion}

The GW detection opens a new window to the FOPT and the Higgs precision measurement is an inevitable path after the Higgs discovery. In this paper, we study their interplay in a specific DM model, namely lepton portal DM model. We emphasize the Higgs portal coupling in this model, which is neglected in the previous literature. The impact is investigated in two aspects. In the cosmological aspect, we have studied the parameter space allowing a FOPT and yielding detectable GW signals at the future detectors, taking LISA as an example. In the particle aspect, we have considered various new channels to further test this model:

\begin{itemize}
\item $pp\to S^+S^-$ at the LHC, which mainly probes $m_S$ and $m_\chi$ since the production is dominated by the Drell-Yan process, can also test the Higgs portal coupling between $h$ and $S^\pm$ (i.e. $\lambda_{HS}$) through the gluon-gluon fusion process.

\item $e^+e^-\to S^{\pm}S^{\mp(*)}$ at the future lepton colliders, which can fill in the gaps between the LEP and LHC constrains ($100~{\rm GeV}\lesssim m_S\lesssim150$ GeV), and probe the $y_\ell$ coupling via the off-shell production of $S^\pm$.

\item Exotic decays of $h$ and $Z$ at the future lepton colliders, which probe the couplings $\lambda_{HS}$ and/or $y_\ell$ for the low $m_\chi$ region.

\item Higgs precision measurements for invisible decay branching ratios and leptonic coupling originated from one-loop contributions, which can provide the best sensitivity for the combination $y_\ell^2\lambda_{HS}$ or $\lambda_{HS}$ assuming $y_\ell$ satisfies the DM
relic abundance requirement.

\item Electron $g-2$ experiments have recently came up with two sets of results. For $\Delta \tilde{a}_e$, DM mass should be larger
than $1$ GeV, while for $\Delta a_e$, DM mass between $0.2\sim2$ GeV is preferred. 
\end{itemize}

In summary, the future Higgs precision measurements can effectively interplay with GW detection, since they both rely on the Higgs portal coupling. The Higgs portal is allowed by this model and can contribute to the Higgs couplings to DM and SM leptons at one-loop level. Therefore, most of the GW detectable parameter space can be cross-checked by the Higgs precision measurement. It shows the rigorous interplay between the future Higgs precision measurement program and the GW detection program.
Specific to the lepton portal DM model, which is hard to probe through DM direct and indirect detections, the Drell-Yan production of charged scalar pair is the useful way to probe this model but only constrains the mass parameter of the scalar and DM.
We studied the Higgs mediated $S^\pm$ pair production, exotic decays of $h/Z$ and electron $g-2$ experiment, which can help extending the constraints on mass parameters and also providing useful constraints on the Yukawa and scalar portal couplings.

\section*{Acknowledgments}
We would like to thank Caterina Doglioni, Manqi Ruan, Jian Wang and Lian-Tao Wang for useful discussions and comments.
The work of JL is supported by National Science Foundation of China under Grant No. 12075005 and by Peking University under startup Grant No. 7101502458. The work of XPW is supported by National Science Foundation of China under Grant No. 12005009. KPX is supported by the Grant Korea NRF-2019R1C1C1010050. KPX would like to thank the hospitality of the University of Chicago where part of this work was performed.

\bibliographystyle{JHEP-2-2.bst}
\bibliography{references}

\end{document}